\documentclass[manuscript]{aastex}

\usepackage{graphics}
\usepackage{amssymb}
\usepackage{longtable}
\usepackage{natbib}

\shorttitle{Adiabatic Mass Loss in Binary Stars. II.}
\shortauthors{Ge et al.}

\begin{document}

\title{Adiabatic Mass Loss in Binary Stars. II. From Zero-Age Main Sequence to the Base of the Giant
Branch}

\author{Hongwei Ge\altaffilmark{1,2}, Ronald F. Webbink\altaffilmark{3}, Xuefei
Chen\altaffilmark{1,2}, and Zhanwen Han\altaffilmark{1,2}}
\altaffiltext{1}{Yunnan Observatories, The Chinese Academy of Sciences, Kunming 650011, China;
hongwei.ge@gmail.com}
\altaffiltext{2}{Key Laboratory for the Structure and Evolution of Celestial Objects, Chinese Academy
of Sciences, Kunming 650011, China}
\altaffiltext{3}{Department of Astronomy, University of Illinois, 1002 W. Green St., Urbana, IL 61801,
USA; rwebbink@illinois.edu}

\begin{abstract}
In the limit of extremely rapid mass transfer, the response of a donor star in an interacting binary
becomes asymptotically one of adiabatic expansion. We survey here adiabatic mass loss from Population
I stars ($Z = 0.02$) of mass $0.10\ M_{\sun}$ to $100\ M_{\sun}$ from the zero age main sequence to
the base of the giant branch, or to central hydrogen exhaustion for lower main sequence stars. The
logarithmic derivatives of radius with respect to mass along adiabatic mass loss sequences translate into
critical mass ratios for runaway (dynamical time scale) mass transfer, evaluated here under the
assumption of conservative mass transfer.  For intermediate- and high-mass stars, dynamical mass
transfer is preceded by an extended phase of thermal time scale mass transfer as the star is stripped of
most of its envelope mass. The critical mass ratio $q_{\rm ad}$\footnote{Throughout this paper, we
follow the convention of defining the binary mass ratio as $q \equiv M_{\rm donor}/M_{\rm
accretor}$.} above which this
\emph{delayed} dynamical instability occurs increases with advancing evolutionary age of the donor
star, by ever-increasing factors for more massive donors. Most intermediate- or high-mass binaries with
nondegenerate accretors probably evolve into contact before manifesting this instability. As they
approach the base of the giant branch, however, and begin developing a convective envelope, $q_{\rm
ad}$ plummets dramatically among intermediate-mass stars, to values of order unity, and a
\emph{prompt} dynamical instability occurs. Among low-mass stars, the prompt instability prevails
throughout main sequence evolution, with $q_{\rm ad}$ declining with decreasing mass, and
asymptotically approaching $q_{\rm ad} = 2/3$, appropriate to a classical isentropic $n = 3/2$ polytrope.
Our calculated $q_{\rm ad}$ agree well with the behavior of time-dependent models by \citet{Che03} of
intermediate-mass stars initiating mass transfer in the Hertzsprung gap. Application of our results to
cataclysmic variables, as systems which must be \emph{stable} against rapid mass transfer, nicely
circumscribes the range in $q_{\rm ad}$ as a function of orbital period in which they are found. These
results are intended to advance the verisimilitude of population synthesis models of close binary
evolution.
\end{abstract}

\keywords{binaries: close --- stars: evolution --- stars: interiors --- stars: mass loss}

\section{Introduction}
\label{intro}

Mass transfer is the defining characteristic distinguishing the evolution of close binary stars from that of
isolated single stars. That mass transfer is typically triggered by the evolutionary expansion of one of the
binary components, but the rate at which mass transfer proceeds depends on the interplay between the
structural response of the donor star to mass loss and the dynamical response of the binary orbit (and
with it, the tidal limit or Roche lobe of the donor star). If the donor star can remain lobe-filling only by
virtue of its evolutionary expansion and/or orbital decay through angular momentum loss, then the donor
star remains in thermal equilibrium, and mass transfer proceeds on that evolutionary expansion/angular
momentum loss time scale. Examples of interacting binaries in this state of slow mass transfer include
classical Algol-type binaries and (most) cataclysmic variables and low-mass X-ray binaries. However, it
is often the case that the donor star's Roche lobe does not expand as rapidly in response to mass loss as
would the donor star itself, if that star were to remain in thermal equilibrium. In this case, the donor will
be driven out of thermal equilibrium. Depending on the thermal structure of the donor's envelope, that
divergence from thermal equilibrium may prevent the donor from expanding far beyond its Roche lobe.
The mass transfer rate is then governed by relaxation of the donor toward thermal equilibrium, i.e., it
proceeds on a thermal time scale. Examples of systems in thermal time scale mass transfer are relatively
rare because of their short lifetimes in mass transfer, but they may include such strongly interacting
binaries as W Serpentis stars \citep{Pla80} and, most prominently, supersoft X-ray sources
\citep{vdH92,Kah97}. In other circumstances, however, thermal relaxation cannot contain expansion of
the donor far beyond its Roche lobe. The mass transfer rate grows inexorably, limited only by
hydrodynamical expansion of the donor envelope through the opening of the Roche potential at the inner
Lagrangian point \citep{Pac72,Sav78,Egg06}, and can in principle approach the mass of the donor star
divided by the orbital period. The prospect of dissipating an appreciable fraction of the donor star's
binding energy on such a short time scale has led to the suggestion that intermediate-luminosity transient
sources \citep[cf.][]{Kas12} are triggered by such dynamical mergers \citep{Mun02,Sok06,Kul07}.

As discussed at some length by \citet[hereafter Paper I]{Ge10a}, and outlined above, the threshold
conditions for dynamical time scale mass transfer depend on the response of the donor star to mass loss,
and on the dynamical response of the orbit and donor Roche lobe to mass transfer, systemic mass loss
and orbital angular momentum loss. Our focus in the present paper is to build model sequences in which
a donor star's specific entropy profile and composition profile are held fixed during mass loss.  These
adiabatic model sequences describe the asymptotic response of donor stars to mass loss in the limit that
the time scales involved are so rapid that thermal relaxation of the donor can be ignored, but not so rapid
that the donor departs in bulk from hydrostatic equilibrium.  At their simplest, previous adiabatic mass
loss models assumed simple polytropic models or variants upon them \citep{Hje87}, or assumed locally
polytropic equations of state \citep{Dai13}. Realistic models with sophisticated equations of state have
been studied by \citet{Hje89a,Hje89b,Ge10a,Ge10b}, and \citet{Del10}, as computing resources have
grown more powerful. We employ here a fully realistic equation of state, as described in Paper I, albeit
retaining the simplification of one-dimensional models.

In this paper, we apply the construction of adiabatic mass-loss sequences, as described in Paper I, to
determining the criteria for dynamical instability in binaries with a radiative donor stars (on the
main-sequence or in the Hertzsprung gap) or with low-mass main-sequence donors. Stars with deep
convective envelopes, that is, those in later evolutionary phases (giant branch and asymptotic giant
branch), respond very differently to mass loss, and present additional issues regarding the  interpretation
of our adiabatic mass loss calculations. We defer discussion of these later evolutionary phases to the next
installment in this series of papers. The present paper is organized as follows. Section~\ref{M-Rdiagr}
introduces the construction of a mass-radius diagram, which provides a useful graphical context in which
we can summarize our results in an immediately-accessible form. Section~\ref{mod-sel} summarizes the
distribution in mass and evolutionary stage (radius) of models chosen to span that diagram. In
Section~\ref{mass-loss} we identify the physical processes that govern the responses of radiative stars to
rapid mass loss in binary systems, and the relationship between prompt and delayed dynamical
instabilities of the donor stars.  Section~\ref{results} presents the results of our survey in both tabular
and graphical form, interpreted in terms of threshold mass ratios
 assuming conservative mass transfer. In Section~\ref{time-dependent} we show that the
threshold mass ratios deduced from our adiabatic mass-loss sequences are qualitatively and quantitatively
consistent with relevant time-dependent mass-loss studies.  An example of the application of these
thresholds to real binary systems, the cataclysmic variables, follows in Section~\ref{appl}. We close
(Section~\ref{discussion}) with a brief summary of results, and a discussion of their limitations.

\section{The Mass-Radius Diagram}
\label{M-Rdiagr}

An efficient vehicle for discussions of interacting binary evolution is the \emph{mass-radius diagram},
illustrating the radii of stars at critical phases of their evolution. Given a companion mass and orbital
separation, this diagram enables one to ascertain immediately the evolutionary stage of the donor star
when it fills its Roche lobe. Furthermore, given some distribution of donor stars in mass and orbital
separation (as proxy for Roche lobe radius), we can see immediately which evolutionary channels are
most frequently populated.

\begin{figure}
\epsscale{.70}
\plotone{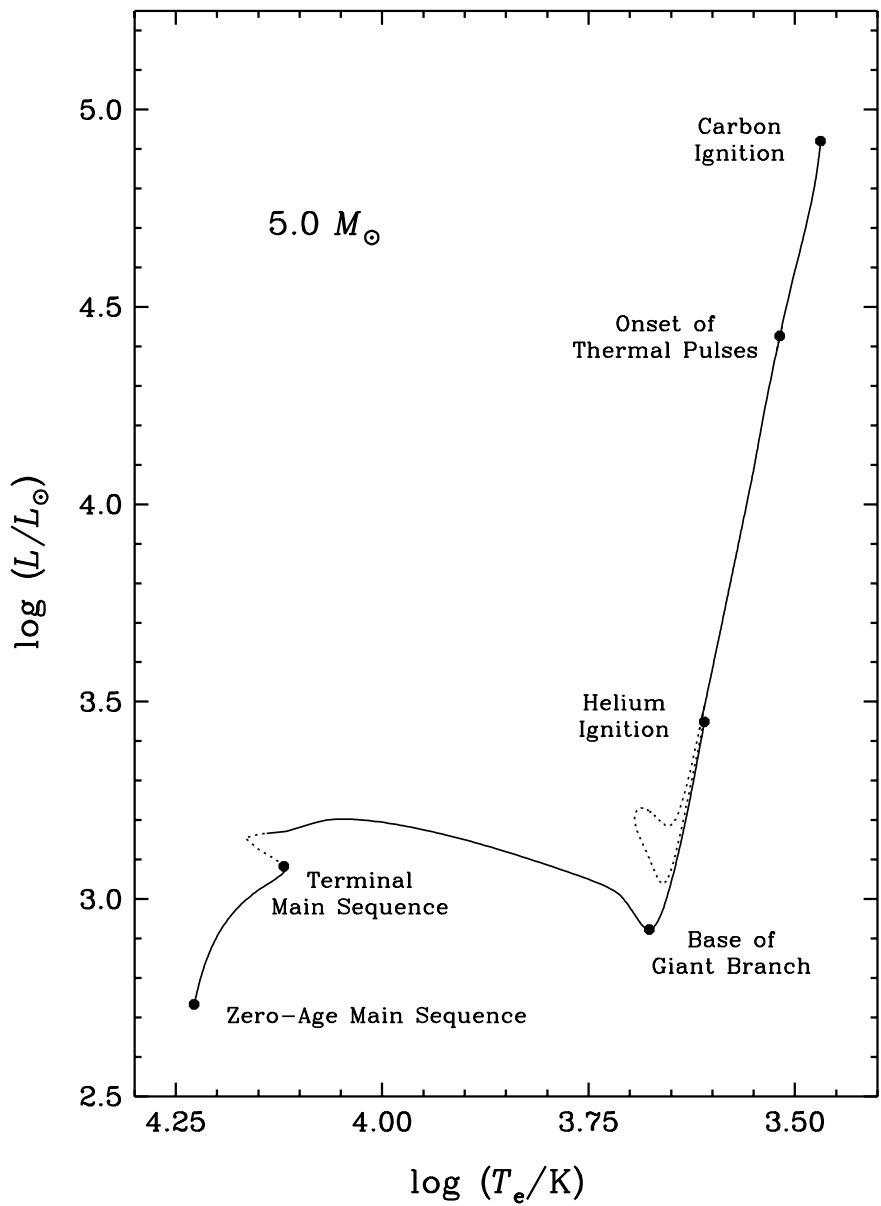}
\caption{The evolutionary track of a $5\ M_{\sun}$ star in the theoretical Hertzsprung-Russell diagram.
Important epochs in its evolution are labeled.  Dotted portions of the evolutionary track signify
evolutionary phases in whih the stellar radius is smaller than in the preceding phase.\label{5Msun_HR}}
\end{figure}

\begin{figure}
\epsscale{.70}
\plotone{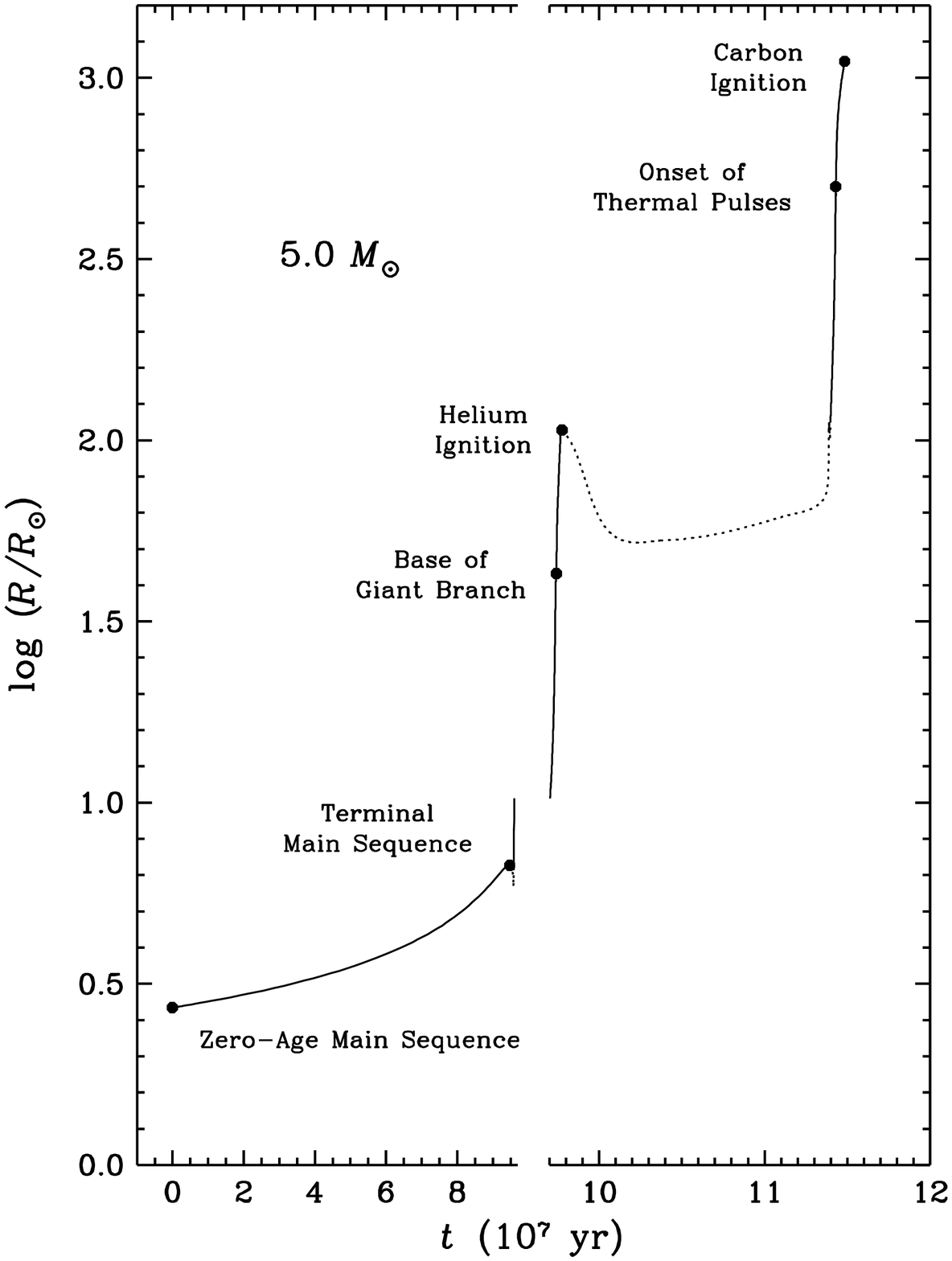}
\caption{Radius of a $5\ M_{\sun}$ star as a function of time.  Fiducial events in its evolution are
labeled. Absent significant angular momentum loss, this star cannot initiate mass transfer during those
phases of its evolution when its radius is smaller than during a preceding phase of evolution (dotted
segments, e.g., during core helium burning, or just beyond the point labeled terminal main sequence).
We refer to these forbidden evolutionary phases as ``shadowed''.\label{5Msun_Rt}}
\end{figure}

\begin{figure}
\plotone{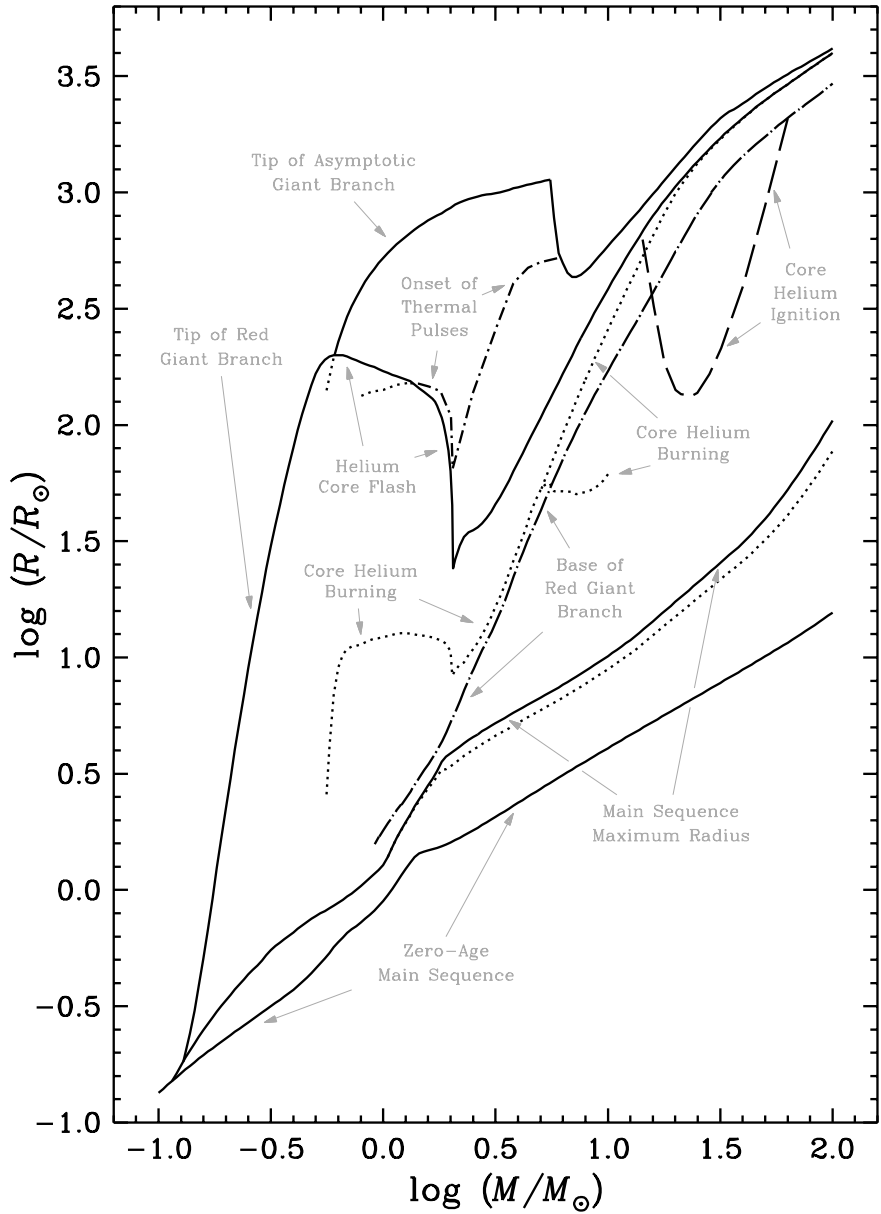}
\caption{The mass-radius diagram, marking fiducial radii as functions of mass.  Solid lines mark the
zero-age main sequence and radius maxima.  Dotted lines mark radius minima, shadowed by preceding
evolutionary phases.  The base of the red giant branch is marked by a dash-dotted line.  Core helium
ignition (where distinguishable from a radius maximum), is marked by a long-dashed line, and the onset
of thermal pulses on the asymptotic giant branch by a short-dashed line.  Stellar wind mass loss has been
neglected throughout.\label{MRdiagr}}
\end{figure}

As an example of the construction of this diagram, consider the evolution of a $5\ M_{\sun}$ star.  Its
evolution in the theoretical Hertzsprung-Russell diagram is shown in Fig.~\ref{5Msun_HR}.  The
position of a star in that diagram immediately fixes its radius, via the blackbody law.  The point in its
evolution at which such a star begins tidal mass transfer then occurs when it first fills its Roche lobe ($R
= R_{\rm L}$).  In the context of close binary evolution, then, the evolution of the donor star radius with
time acquires special significance, as shown in Fig.~\ref{5Msun_Rt}.  In the context of our $5\
M_{\sun}$ example, we see that if $0.434 \le \log (R_{\rm L}/R_{\sun}) < 0.827$ it first fills its Roche
lobe during core hydrogen burning, while still on the main sequence. If $0.827 \le \log (R_{\rm
L}/R_{\sun}) < 2.028$, it does so as it crosses the Hertzsprung gap, or during its initial ascent of the
giant branch, prior to core helium burning; and if $2.028 \le \log (R_{\rm L}/R_{\sun}) < 3.046$ during
ascent of the asymptotic giant branch. But whenever an evolving star spontaneously contracts, it cannot
ordinarily initiate mass transfer, as it will have done so during a prior phase of
evolution.\footnote{Exceptions can occur, for example, through encounters with field stars, or orbital
variations driven by a more distant companion in a triple system \citep{Koz62}.  For this reason, our
survey of adiabatic responses encompasses all phases of evolutionary expansion, including those
shadowed by prior evolution.} In the example at hand, the $5\ M_{\sun}$ star will not fill its Roche
lobe during its momentary contraction at the terminal main sequence, or during core helium burning until
it reaches a radius on the asymptotic giant branch equal to its prior radius at helium ignition. We refer to
these excluded phases of contraction as shadowed by one or more prior phases of evolution. Identifying
critical radii of stars in similar fashion throughout our library of evolutionary models, we can construct a
diagram of these critical radii as a function of mass (Fig.~\ref{MRdiagr}).

\begin{figure}
\plotone{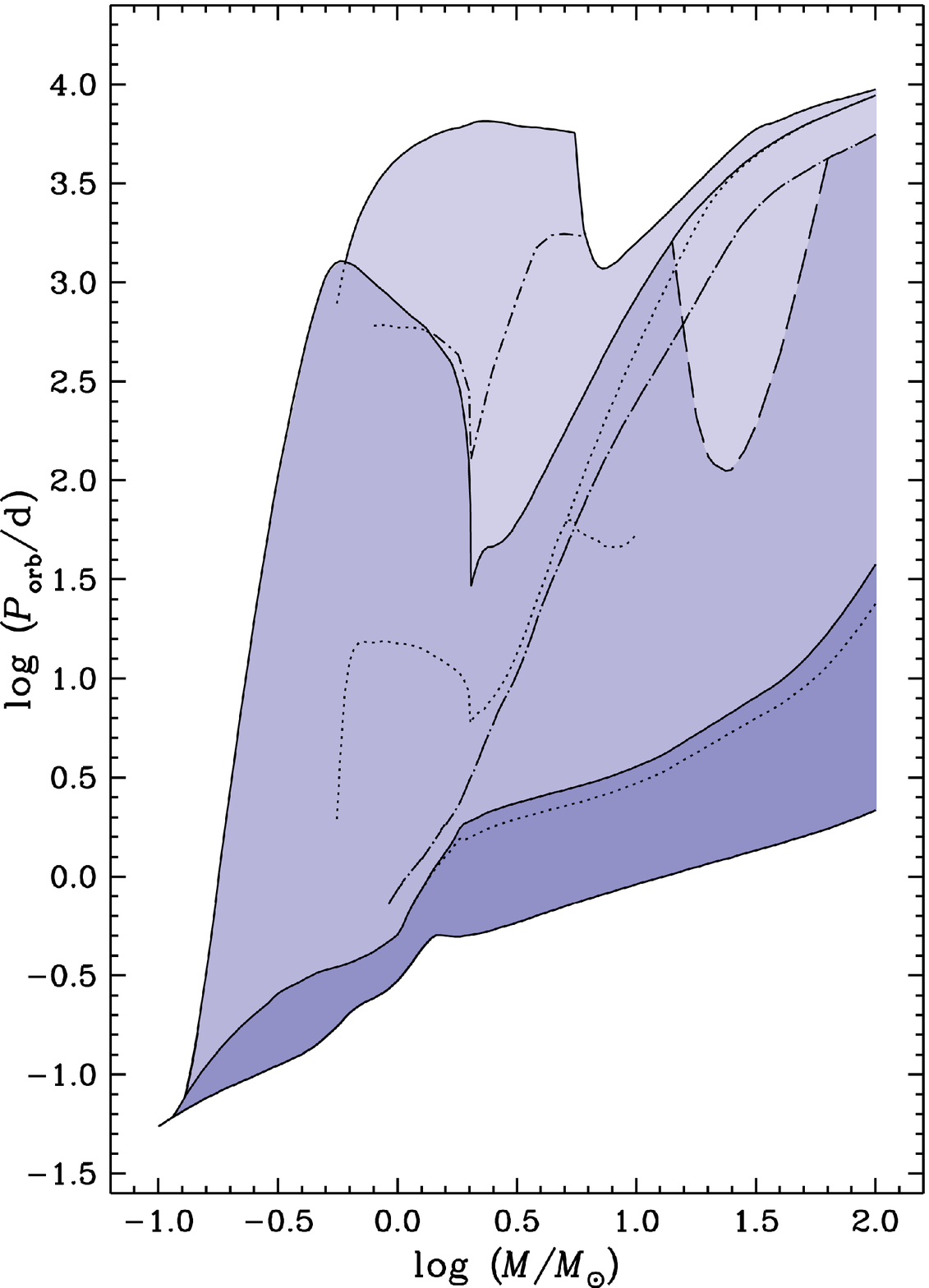}
\caption{The mass-orbital period diagram corresponding to Fig.~\ref{MRdiagr}.  Mass ratio $q = 1$ has
been assumed.  Line segments are coded as in Fig.~\ref{MRdiagr}.  Background coloring reflects the
historical classification of modes of mass transfer according to the evolutionary state of the donor star at
the onset of mass transfer \citep{Kip67,Lau69}: Case A (central hydrogen burning) in dark blue, Case
B (shell hydrogen burning leading to helium ignition) in medium blue, and Case C (expansion
post-helium ignition) in light blue, respectively.
\label{MPdiagr}}
\end{figure}

The mass-radius diagram is directly related to a mass-orbital period diagram (Fig.~\ref{MPdiagr}),
useful in identifying which mass transfer channels observed binaries may follow. Using the
\citet{Egg83} approximation for the dimensionless Roche lobe radius, $r_{\rm L} = R_{\rm L}/A$, with
$A$ the orbital separation, we have
\[
\log (P_{\rm orb}/{\rm d}) = \frac{3}{2} \log (R_L/R_{\sun}) - \frac{1}{2} \log (M/M_{\sun}) + \log
g(q) - 0.45423 \ ,
\]
where $g(q)$ is a very weak function of $q$, the ratio of donor to accretor mass:
\[
g(q) = \left( \frac{2q}{1+q} \right)^{1/2} \left( \frac{0.6 + q^{-2/3} \ln (1 + q^{1/3})}{0.6 + \ln 2}
\right)^{3/2}
\]
($g(1)=1$, by construction).  Thus, the orbital period of a binary with donor mass $M$ fixes (to within a
weak function of the mass ratio) the radius of that prospective donor at which it fills its Roche lobe ($R =
R_{\rm L}$).

\section{Model Selection}
\label{mod-sel}

The evolutionary code employed for this study was based on the stellar evolution code developed by
\citet{Egg71,Egg72,Egg73} and \citet{Pax04}.  It is a one-dimensional (spherically symmetric)
non-Lagrangian code, and includes a treatment of convective overshooting as described by \citet{SPE97}
with overshooting parameter $\delta_{\rm ov}$.  A more detailed account of the physics it incorporates
can be found in Section 2.2 of Paper I.

\begin{figure}
\plotone{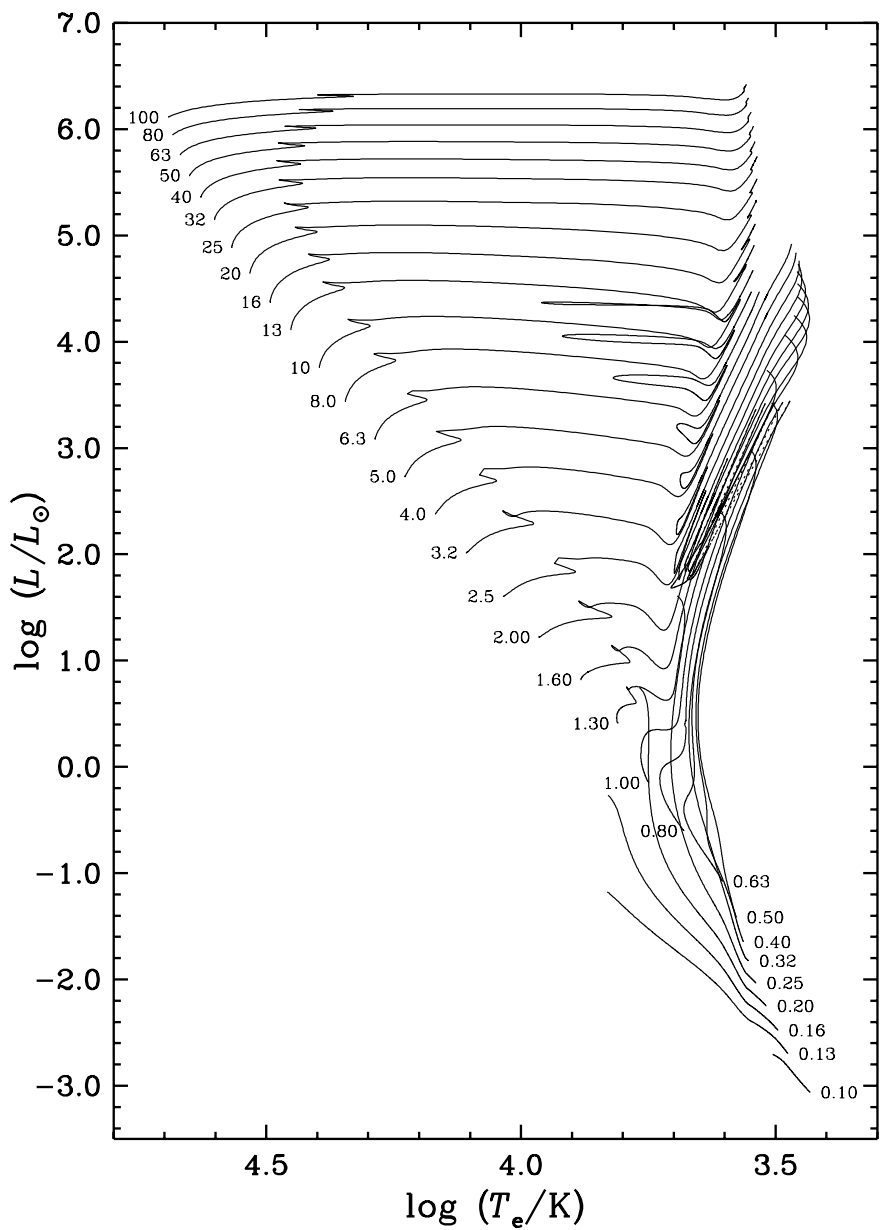}
\caption{The Hertzsprung-Russell diagram for stellar models included in this survey.  Evolutionary
tracks are labeled by mass (in solar units).  Evolution beyond the maximum radius for each mass has
been omitted.  Some masses (0.22, 0.28, 0.36, 0.45, 0.56, 0.71, 0.89, 1.14, 1.439, 1.80, and 2.04
$M_\sun$) have been omitted for clarity.\label{HRdiagr}}
\end{figure}

The initial models for the mass loss sequences reported in this series of papers were selected from a
library of stellar evolution sequences of nominal Population I metallicity ($Z = 0.02$), spanning the full
range of normal stellar masses (see Fig.~\ref{HRdiagr}). We assume evolution at constant mass up to the
onset of tidal mass transfer, as this establishes a definitive reference point in the absence of an a priori
physical theory to quantify mass loss. (Otherwise, one needs to introduce an empirical mass-loss
prescription that invariably involves introducing additional, empirical, model parameters.) This
assumption is clearly inappropriate to the most massive and most luminous stars considered here, but to
the extent that these stars are roughly in thermal equilibrium as they reach their Roche lobes, their
response to mass loss depends only on their instantaneous mass and composition profile, without regard
to prior mass loss history. Our mass loss models should therefore capture the most important physical
processes at play.

Altogether, 42 evolutionary sequences formed the basis of this study.  They were selected at intervals of
$\Delta \log M \approx 0.1$ over the range $-1 \le \log (M/M_{\sun}) \le 2$, with additional models at
intervals of $\Delta \log M \approx 0.05$ in the interval $-0.7 < \log (M/M_{\sun}) < 0.3$, and one
additional sequence at $M = 2.04\;M_{\sun}$ marking the transition from degenerate to non-degenerate
helium ignition.  Models consisted of 199 to 1299 mesh points, depending on the complexity of their
structure at advanced phases of evolution, with typically of order $10^3$ models in each evolutionary
sequence.

\begin{figure}
\plotone{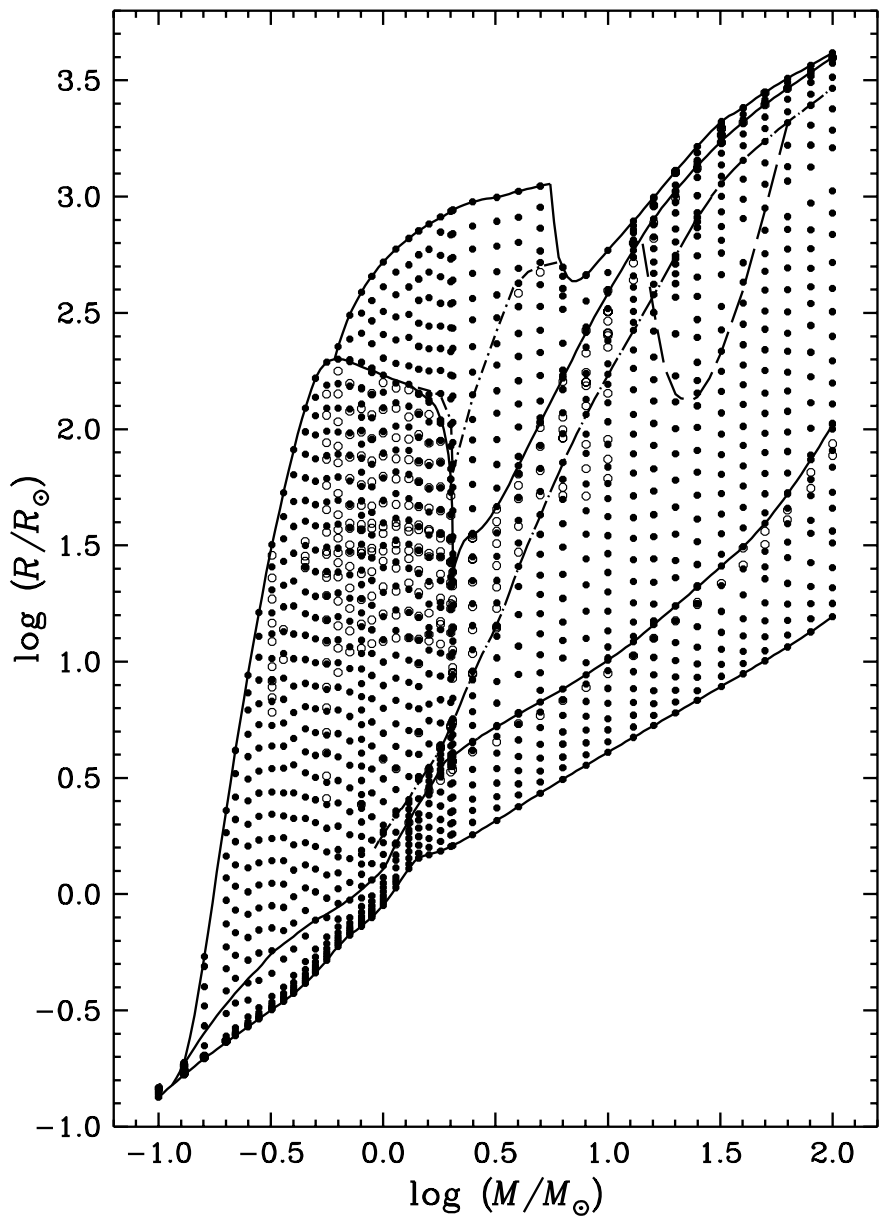}
\caption{The distribution in the mass-radius diagram of initial models for adiabatic mass-loss sequences:
Circles mark the selected models, filled ($\bullet$) if unshadowed, open ($\circ$) if shadowed.  Only
models that have not evolved beyond the base of the giant branch are included in the present
study.\label{models}}
\end{figure}

For each of these evolutionary sequences, initial models for adiabatic mass loss sequences were selected
to coincide with evolutionary extrema in radius, starting from the zero-age main sequence.  Additional
mass loss sequences were constructed at intervals of $\Delta \log R \approx 0.1$ during all phases of
evolutionary expansion, including those shadowed by prior evolution (as these might still be relevant
in dense stellar environments or in multiple star systems).  The main sequence, up to central hydrogen
exhaustion, was sampled at intervals $\Delta X_{\rm c} \approx 0.1$ in central hydrogen abundance.  All
told, our library numbers 1670 adiabatic mass loss sequences (Fig.~\ref{models}), typically of order
$10^3$ models per mass loss sequence.  Of these sequences, 680 fall within the scope of this paper
(evolution to the base of the giant branch), with the balance to be presented in the next installment.

In this paper, we present results for a subset of these model sequences, covering evolutionary phases
from zero-age main sequence, through central hydrogen exhaustion, up to arrival at the base of the giant
branch, which we take to coincide with the luminosity minimum seen in the evolutionary tracks of
intermediate-mass and massive stars (see Fig.~\ref{HRdiagr}). This luminosity minimum does not exist
for low-mass stars; we include only up through central hydrogen exhaustion for these low-mass stars.
Tables~\ref{intmod} and~\ref{glbmod} document the initial properties of the donor stars at the
beginning of each mass-loss sequence.

\begin{deluxetable}{rrrrrrrrrrr}

\tabletypesize{\footnotesize}
\tablewidth{0pt}
\tablecolumns{11}
\tablecaption{Interior properties of initial models\label{intmod}}

\tablehead{
\colhead{$k$} & \colhead{$t$} & \colhead{$M_{\rm ce}$} & \colhead{$M_{\rm c}$} &
\colhead{$M_{\rm ic}$} & \colhead{$\psi_{\rm c}$} & \colhead{$\log \rho_{\rm c}$} &
\colhead{$\log
T_{\rm c}$} & \colhead{$X_{\rm c}$} & \colhead{$Y_{\rm c}$} & \colhead{$X_{\rm s}$} \\
 \colhead{} & \colhead{yr} & \colhead{$M_\sun$} & \colhead{$M_\sun$} & \colhead{$M_\sun$} &
\colhead{} &
\colhead{$\rm g\:cm^{-3}$} & \colhead{K} & \colhead{} & \colhead{} & \colhead{}
}
\startdata
\cutinhead{\normalsize $5.0000\ M_\sun$}
1 & 2.232372E+04 & 0.0000 & 1.2380 & 0.0000 & --4.227 & 1.277 & 7.423 & 0.700 & 0.280 & 0.700
\\*
2 & 2.589656E+07 & 0.0000 & 1.5601 & 0.0000 & --4.340 & 1.265 & 7.430 & 0.602 & 0.379 & 0.700
\\
3 & 4.490234E+07 & 0.0000 & 1.4860 & 0.0000 & --4.436 & 1.262 & 7.439 & 0.509 & 0.472 & 0.700
\\
4 & 6.151553E+07 & 0.0000 & 1.4051 & 0.0000 & --4.533 & 1.268 & 7.450 & 0.405 & 0.576 & 0.700
\\
5 & 7.390986E+07 & 0.0000 & 1.3283 & 0.0000 & --4.614 & 1.283 & 7.462 & 0.305 & 0.676 & 0.700
\\
6 & 8.414729E+07 & 0.0000 & 1.2465 & 0.0000 & --4.683 & 1.314 & 7.478 & 0.197 & 0.783 & 0.700
\\
7 & 9.095053E+07 & 0.0000 & 1.1770 & 0.0000 & --4.714 & 1.366 & 7.498 & 0.106 & 0.875 & 0.700
\\
8 & 9.572705E+07 & 0.0000 & 1.1155 & 0.0000 & --4.656 & 1.486 & 7.539 & 0.023 & 0.957 & 0.700
\\
9 & 9.680205E+07 & 0.0000 & 1.0988 & 0.0000 & --4.260 & 1.800 & 7.627 & 0.000 & 0.980 & 0.700
\enddata

\tablecomments{Table \ref{intmod} is published in its entirety in the electronic edition of the {\it
Astrophysical Journal}.  A portion is shown here for guidance regarding its form and content.}

\end{deluxetable}

\newcounter{tbl1}
Table~\ref{intmod} is arranged in segments, by stellar mass, $M_i$.  Successive columns list:
\begin{list}{(\arabic{tbl1})}{\usecounter{tbl1}}
\item $k$ --- mass loss sequence number;
\item $t$ --- age (yr);
\item $M_{\rm ce}$ --- mass of the convective envelope ($M_\sun$);
\item $M_{\rm c}$ --- core mass ($M_\sun$);
\item $M_{\rm ic}$ --- inner core mass ($M_\sun$);
\item $\psi_{\rm c}$ --- central electron chemical potential ($\mu_e$, in units of $kT$);
\item $\log \rho_{\rm c}$ --- central density (${\rm g\,cm^{-3}}$);
\item $\log T_{\rm c}$ --- central temperature (K);
\item $X_{\rm c}$ --- central hydrogen abundance (fraction by mass);
\item $Y_{\rm c}$ --- central helium abundance (fraction by mass); and
\item $X_{\rm s}$ --- surface hydrogen abundance (fraction by mass)
\end{list}

Age $t$ is measured from the zero-age main sequence model (excluding pre-main-sequence evolution).
The mass of the convective envelope $M_{\rm ce}$ refers to the mass depth of the base of the outermost
convection zone. The core mass $M_{\rm c}$ refers to the mass coordinate at which the helium
abundance is halfway between the surface helium abundance and the maximum helium abundance in the
stellar interior. The inner core mass $M_{\rm ic}$ identifies the mass coordinate at which the helium
abundance is halfway between the maximum helium abundance in the stellar interior and the minimum
helium abundance interior to that maximum; in the absence of measurable helium depletion in the
hydrogen-exhaused core, $M_{\rm ic}$ is set to a default value of zero.  $M_{\rm c}$ and $M_{\rm ic}$
characterize the \emph{range} in mass over which hydrogen and helium are being depleted during their
respective core burning phases, and \emph{not} the amount of mass that has been consumed.  Upon core
fuel exhaustion, $M_{\rm c}$ and $M_{\rm ic}$ mark the midpoints in hydrogen and helium depletion
profiles, respectively.  The dimensionless central electron chemical potential $\psi_{\rm c}$ measures
the degree of electron degeneracy ($\psi_{\rm c} > 0$).

\begin{deluxetable}{rrrrrrrrrr}

\tabletypesize{\footnotesize}
\tablewidth{0pt}
\tablecolumns{10}
\tablecaption{Global properties of initial models\label{glbmod}}

\tablehead{
\colhead{$k$} & \colhead{$\log R$} & \colhead{$\log T_{\rm e}$} & \colhead{$\log L$} &
\colhead{$\log L_{\rm H}$} & \colhead{$\log L_{\rm He}$} & \colhead{$\log L_Z$} & \colhead{$\log
|L_\nu|$} & \colhead{$\log |L_{\rm th}|$} & \colhead{$I/MR^2$} \\
\colhead{} & \colhead{$R_\sun$} & \colhead{K} & \colhead{$L_\sun$} & \colhead{$L_\sun$} &
\colhead{$L_\sun$} & \colhead{$L_\sun$} & \colhead{$L_\sun$} & \colhead{$L_\sun$} & \colhead{}
}
\startdata
\cutinhead{\normalsize $5.0000\ M_\sun$}
1 & 0.4342 & 4.2276 & 2.7323 & 2.786 & --23.170 & \nodata\phn & --5.182* & 0.443* & 0.0606 \\*
2 & 0.4819 & 4.2201 & 2.7975 & 2.849 & --22.757 & \nodata\phn & --5.130* & --0.767* & 0.0551 \\
3 & 0.5288 & 4.2111 & 2.8552 & 2.906 & --22.361 & \nodata\phn & --5.079* & --0.955* & 0.0502 \\
4 & 0.5854 & 4.1976 & 2.9144 & 2.965 & --21.900 & \nodata\phn & --5.023* & --1.159* & 0.0451 \\
5 & 0.6452 & 4.1804 & 2.9652 & 3.016 & --21.417 & \nodata\phn & --4.970* & --1.184* & 0.0407 \\
6 & 0.7165 & 4.1565 & 3.0124 & 3.062 & --20.807 & \nodata\phn & --4.905* & --1.009* & 0.0363 \\
7 & 0.7828 & 4.1321 & 3.0471 & 3.096 & --20.077 & \nodata\phn & --4.821* & --0.823\phn & 0.0330
\\
8 & 0.8270 & 4.1190 & 3.0832 & 3.130 & --18.596 & \nodata\phn & --4.606* & 0.515\phn & 0.0304 \\
9 & 0.7685 & 4.1662 & 3.1551 & 3.134 & --15.512 & \nodata\phn & --4.046* & 2.225\phn & 0.0296
\enddata

\tablecomments{Table \ref{glbmod} is published in its entirety in the electronic edition of the {\it
Astrophysical Journal}.  A portion is shown here for guidance regarding its form and content.}

\end{deluxetable}

\newcounter{tbl2}
Like Table~\ref{intmod}, Table~\ref{glbmod} is arranged in segments, by stellar mass, $M_i$.
Successive columns list:
\begin{list}{(\arabic{tbl2})}{\usecounter{tbl2}}
\item $k$ --- mass loss sequence number;
\item $\log R$ --- radius ($R_\sun$);
\item $\log T_e$ --- effective temperature (K);
\item $\log L$ --- stellar luminosity ($L_\sun$);
\item $\log L_{\rm H}$ --- hydrogen-burning luminosity ($L_\sun$);
\item $\log L_{\rm He}$ --- helium-burning luminosity ($L_\sun$);
\item $\log L_Z$ --- heavy-element (carbon-, oxygen-, etc.) burning luminosity ($L_\sun$);
\item $\log |L_{\nu}|$ --- log neutrino luminosity ($L_\sun$, with asterisk, *, appended to signify that
this
is a \emph{negative} contribution to the net stellar luminosity);
\item $\log |L_{\rm th}|$ --- gravothermal luminosity ($L_\sun$, with asterisk, *, appended where the
gravothermal luminosity is negative); and
\item $I/RM^2$ --- dimensionless moment of inertia
\end{list}

\section{Adiabatic Mass Loss}
\label{mass-loss}

\subsection{Mass Outflow Near $L_1$}
\label{outflow}

The mass transfer rate in a binary system is determined by fluid flow through the region around the inner
Lagrangian point, $L_1$, where Roche potentials open to the companion star. Far from the inner
Lagrangian point, the donor star departs negligibly from hydrostatic equilibrium, as shown by
\citet{Pac72} and \citet{Egg06}, so long as the radial excess beyond its Roche lobe is small. This allows
us to adopt a semi-physical 1D model (described in the Appendix to Paper I) relating the mass loss rate
from the donor star in question to the structure of its envelope beyond the inner critical surface far from
the inner Lagrangian point. In reality, this is at best a rough approximation. It assumes laminar flow
along equipotential surfaces, with the specific enthalpy of the flow along any streamline described in
terms of the pressure, density, and adiabatic exponents at the source of that streamline. Real streamlines
will inevitably be broken up by turbulence and Coriolis effects (which impede mass loss), and readily
cross equipotential surfaces. In radiative envelopes, stable density stratification (negative buoyancy)
inhibits upwelling from the stellar interior, so mass flow toward the inner Lagrangian region can be
expected to be dominated by surface flows. In contrast, mass loss from convectively unstable envelopes
can be expected to be dominated by upwelling along the line of centers, and may exceed our 1D
estimates by some unknown factor.

It follows from the preceding discussion that the onset of dynamical time scale mass transfer cannot be
instantaneous, but accelerates from an initial trickle to full-blown dynamical instability as it surpasses the
stellar thermal time scale rate, i.e., as the flow asymptotically becomes adiabatic. Accordingly, we define
the onset of dynamical time scale mass transfer not from the instant a donor star fills its Roche lobe, but
rather from the instant at which the Roche lobe penetrates deeply enough into the stellar envelope to
drive thermal time scale mass transfer. A detailed account of this formalism may be found in Paper I. We
calculate the critical mass ratio for dynamical instability as the limiting mass ratio for which the adiabatic
mass loss sequence just reaches thermal time scale mass transfer.

Note that this estimate assumes that we can reasonably approximate the structure of the donor star
envelope beyond its Roche lobe using adiabatic mass loss models. In reality, the outer envelope of a star
relaxes to thermal equilibrium much faster than does the star as a whole. For stars with radiative
envelopes, this thermal relaxation is characterized by absorption of energy from the radiation field, so the
overflow layers will have higher specific entropy than modeled by pure adiabatic expansion. We should
therefore expect that radiative stars drive higher mass transfer rates than our adiabatic mass loss models
predict, and therefore a shallower degree of overflow is needed to drive thermal time scale mass loss.
Accordingly, our critical mass ratios for radiative stars are likely systematically \emph{overestimated}.
That is, thermal relaxation within a radiative envelope tends to make a star \emph{more} unstable against
rapid mass transfer. However, because specific entropy varies extremely rapidly with mass in the outer
envelopes of radiative stars, we expect this effect to be small, as evidenced in
Section~\ref{time-dependent} below.

In contrast, thermal relaxation in the outer envelopes of convective stars tends to depress specific entropy
near the surface; energy may be lost to the radiation field of the star, whence it is radiated from the stellar
photosphere. We see the result in the rapid superadiabatic expansion of the surface layers in adiabatic
mass loss sequences calculated using standard mixing-length models for initial models. At face value,
this excessive expansion would suggest that our algorithm for finding the critical mass ratio for
dynamical time scale mass transfer exaggerates the tendency toward runaway mass transfer, and so
\emph{underestimates} the critical mass ratio.

As emphasized by \citet{Woo11}, for example, thermal relaxation within the superadiabatic outer layers
of a surface convection zone is extremely rapid, even for dynamical outflow rates. To a first
approximation, the entropy profile of that superadiabatic region migrates homologously inward in step
with mass loss, as can be seen in Figure~2 of \citet{Woo11}. Clearly, our adiabatic mass loss sequences
suppress this
thermal relaxation, but we can mimic it for purposes of evaluating critical mass ratios by constructing
artificial mass loss sequences in which the outer convection zone is replaced by a completely isentropic
envelope, with specific entropy fixed at the base of that convection zone. The construction of these
artificially isentropic envelope models is described in more detail in Paper I, where they were termed
\emph{pseudo}-models. Their initial radii, $\tilde{R}_i$ are inflated with respect to more realistic
(mixing-length) models (of radii $R_i$); we characterize the degree of inflation by the parameter
$\Delta_{\rm exp} \equiv \log (\tilde{R}_i/R_i)$. Our premise, then, in constructing artificially
isentropic envelope mass loss sequences is that their outer entropy profiles migrate homologously inward
with mass loss, as do the profiles of realistic models with thermal relaxation, and so the artificial
sequence closely parallels a time-dependent sequence, but with nominal radii inflated by a factor
$10^{\Delta_{\rm exp}}$. We therefore consider the threshold mass-radius exponent and corresponding
limiting mass ratio for conservative mass transfer as derived from the artificially isentropic envelope
models ($\tilde{\zeta}_{\rm ad} \equiv (\partial \ln \tilde{R}/\partial \ln M)_{\rm ad}$ and
$\tilde{q}_{\rm ad}$, respectively) more realistic than those derived from adiabatic mass loss sequences
for standard mixing-length models ($\zeta_{\rm ad} \equiv (\partial \ln R/\partial \ln M)_{\rm ad}$ and
$q_{\rm ad}$). For the low-mass main sequence stars included in this paper, convection is generally
quite efficient, even near the stellar surface, and so the difference between mixing-length envelopes and
isentropic ones is minimal. But we shall see in the next paper in this series that this is not necessarily the
case for giant branch and asymptotic giant branch stars.

\subsection{Structural Response of Radiative Stars}
\label{response}

Stars with very shallow or nonexistent surface convection zones contract rapidly in response to adiabatic
mass loss. This response is a consequence of several factors. Radiative stars generally have much more
centrally-condensed mass distributions than do stars with deep surface convection zones, so a given
decrement of mass loss removes a proportionately larger volume of material from radiative envelopes
than from convective envelopes. Within radiative envelopes, stellar opacity is typically dominated by
free-free and bound-free absorption. These opacities are Kramers-like ($\kappa \propto \rho T^{-7/2}$),
and so increase rapidly with decreasing temperature and pressure, as the envelope is decompressed. The
radiative flux through the envelope is therefore choked off, the surface luminosity and stellar radius
decrease precipitously, and the surface density increases rapidly, accompanied by a relatively modest
decrease in surface temperature.

\begin{figure}
\includegraphics[scale=0.70]{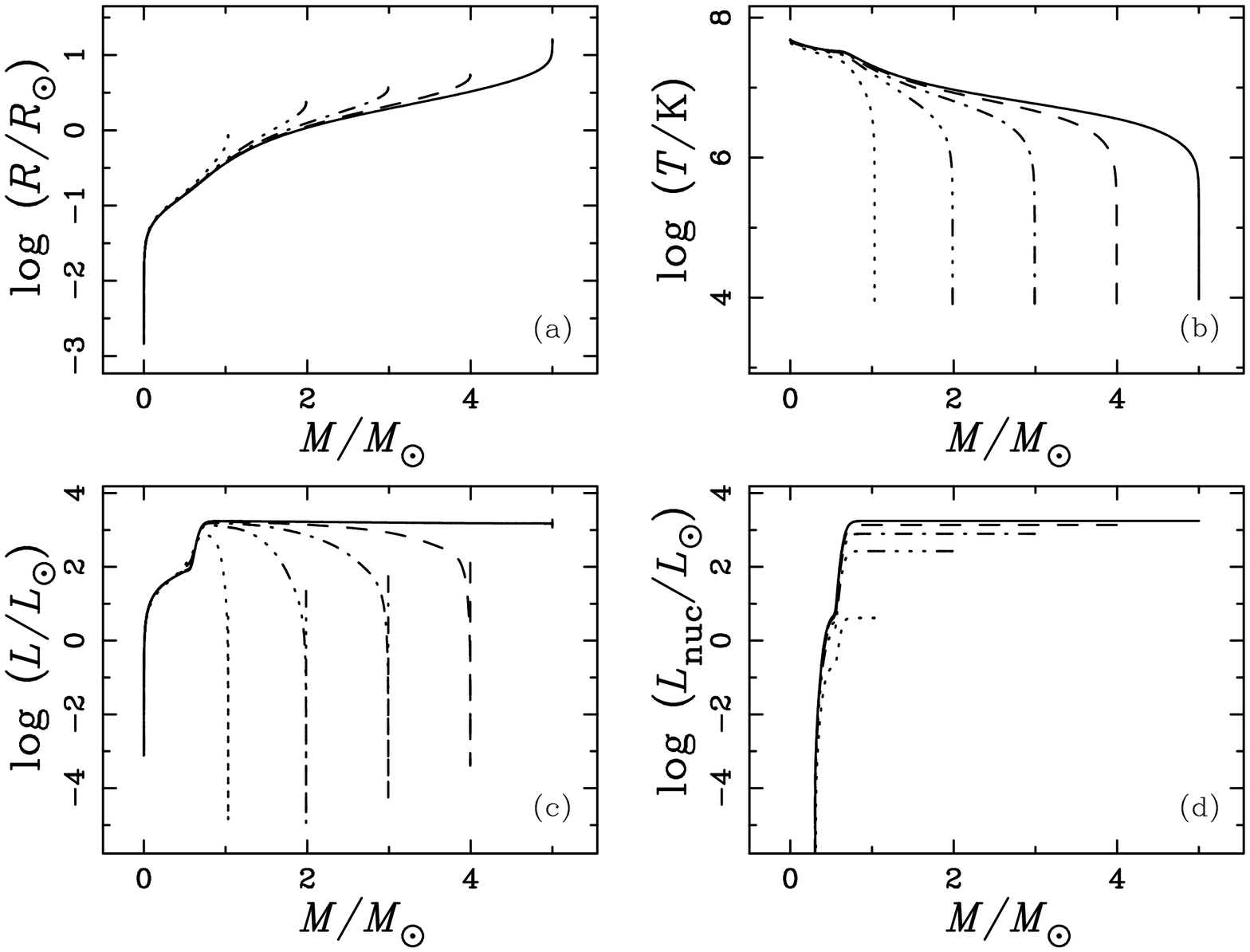}
\caption{Adiabatic response to mass loss of a $5\ M_\sun$ star in the Hertzsprung gap (sequence $k =
14$ in Tables~\ref{intmod}-\ref{dynml} below).  Snapshot interior profiles as functions of remaining
mass (5, 4, 3, 2, and $1\ M_\sun$) are shown for (a) radius), (b) temperature, (c) luminosity, and (d)
nuclear luminosity.  The difference between local luminosity (c) and interior nuclear luminosity (d)
reflects energy absorption (or release) by the decompressed stellar envelope.\label{5M_Mid_HG}}
\end{figure}

This chain of events is illustrated in Fig.~\ref{5M_Mid_HG} for a $5\ M_{\sun}$ star midway in
crossing the Hertzsprung gap. This star has exhausted hydrogen in a non-degenerate core surrounded by a
thick hydrogen-burning shell centered at mass $1.101\ M_{\sun}$. It is expanding rapidly toward the
giant branch ($R/\dot{R} = 2.83 \times 10^5\ {\rm yr}$, compared to a thermal time scale of order
$GM^2/RL = 3.24 \times 10^4\ {\rm yr}$).  Roughly 9\% of its nuclear luminosity is absorbed in driving
this expansion. Near the surface of this star, the scale height for thermodynamical variables (density,
temperature, pressure, entropy) becomes extremely small, so removal of the outermost mass layers results
in a rapid decrease in specific entropy at the stellar surface, and with it the rapid increase in density
described above, leading to the precipitous initial decrease in radius seen in Fig.~\ref{5M_Mid_HG}a.
This decrease is so rapid that the donor star is initially stable against dynamical time scale mass transfer
for any mass ratio of interest. Roche lobe overflow may then be driven by thermal relaxation of the donor
star, or by its evolutionary expansion. However, as mass loss proceeds, surface entropy gradients (which
are fixed in mass by the adiabatic assumption) become shallower, and the rapid contraction in stellar
radius moderates in the adiabatic limit. If the binary mass ratio is high enough ($M_{\rm donor}/M_{\rm
accretor} > 4.73$ for conservative mass transfer), this donor star may develop a delayed dynamical
instability, as described in Paper I. As seen in Fig.~\ref{5M_Mid_HG}b, the temperature profile marches
inward as mass loss proceeds with relatively little change in shape until the star is nearly stripped to its
helium core. That profile is tied closely to the pressure profile (not shown), and reflects the rapid
decrease in pressure scale height near the instantaneous stellar surface as mass loss proceeds.
Fig.~\ref{5M_Mid_HG}c shows the dramatic drop in stellar luminosity that results from the rapid
increase in Kamers-like opacity under decompression, as described above. In contrast, the nuclear
luminosity (Fig.~\ref{5M_Mid_HG}d) is affected relatively little until the stellar mass approaches the
hydrogen-burning shell.

\section{Results}
\label{results}

Table~\ref{dynml} summarizes the quantitative results of our investigation for both those model
sequences derived from initial models with standard mixing-length convective envelopes (columns 2-7),
and those sequences derived from initial models with artificially isentropic convective envelopes
(columns 8-13).  For each set of sequences, it identifies critical points marking the onset of runaway
(dynamical time scale) mass transfer, and the (critical) initial conditions (mass-radius exponent and mass
ratio) corresponding to those critical points. As noted in Section~\ref{outflow}, the onset of dynamical
time scale mass transfer is not instantaneous, but is preceded by an episode of accelerating mass transfer.
We associate the transition to dynamical time scale mass transfer with the mass transfer rate equaling the
nominal thermal time scale rate of the initial model of the sequence, $\dot{M}_{\rm KH} = - R_i
L_i/GM_i$; beyond that rate, the response of the donor becomes asymptotically adiabatic. We then
define the critical mass ratio for dynamical mass transfer as the minimum initial mass ratio for which
$\dot{M}$ reaches $\dot{M}_{\rm KH}$.

\begin{deluxetable}{rrrrrrrrrrrrrr}

\tabletypesize{\footnotesize}
\rotate
\tablewidth{0pt}
\tablecolumns{14}
\tablecaption{Thresholds for conservative dynamical time scale mass transfer\label{dynml}}

\tablehead{
\colhead{} & \multicolumn{6}{c}{Mixing-length convection} & \colhead{} &
\multicolumn{6}{c}{Isentropic convection} \\[8pt]
\cline{2-7} \cline{9-14}
\colhead{$k$} & \colhead{$\log R_i$} & \colhead{$M_{\rm KH}$} & \colhead{$\log R_{\rm
KH}$} & \colhead{$\log R^*_{\rm KH}$} & \colhead{$\zeta_{\rm ad}$} & \colhead{$q_{\rm ad}$} &
& \colhead{$\Delta_{\rm exp}$} & \colhead{$\tilde{M}_{\rm KH}$} & \colhead{$\log
\tilde{R}_{\rm KH}$} & \colhead{$\log \tilde{R}^*_{\rm KH}$} & \colhead{$\tilde{\zeta}_{\rm ad}$}
& \colhead{$\tilde{q}_{\rm ad}$} \\
\colhead{} & \colhead{$R_\sun$} & \colhead{$M_\sun$} & \colhead{$R_\sun$} & \colhead{$R_\sun$}
& \colhead{} & \colhead{} & \colhead{} & \colhead{} & \colhead{$M_\sun$} & \colhead{$R_\sun$} &
\colhead{$R_\sun$} & \colhead{} & \colhead{}
}
\startdata
\cutinhead{\normalsize $5.0000\ M_\sun$}
 1 & 0.4342 & 2.9669 & 0.1873 & 0.1885 & 3.406 & 2.373 & & 0.0003 & 2.9662 & 0.1873 & 0.1885 &
3.409 &
2.374 \\*
2 & 0.4819 & 3.0121 & 0.2048 & 0.2062 & 3.739 & 2.529 & & 0.0002 & 3.0115 & 0.2048 & 0.2062 &
3.741 & 2.530 \\
3 & 0.5289 & 3.0535 & 0.2236 & 0.2252 & 4.069 & 2.684 & & 0.0002 & 3.0529 & 0.2236 & 0.2252 &
4.072 & 2.685 \\
4 & 0.5855 & 3.0987 & 0.2488 & 0.2505 & 4.463 & 2.868 & & 0.0003 & 3.0981 & 0.2487 & 0.2504 &
4.467 & 2.870 \\
5 & 0.6453 & 3.1432 & 0.2782 & 0.2801 & 4.869 & 3.059 & & 0.0003 & 3.1426 & 0.2782 & 0.2801 &
4.873 & 3.060 \\
6 & 0.7166 & 3.1932 & 0.3169 & 0.3191 & 5.337 & 3.278 & & 0.0003 & 3.1925 & 0.3168 & 0.3191 &
5.342 & 3.281 \\
7 & 0.7829 & 3.2385 & 0.3557 & 0.3582 & 5.763 & 3.478 & & 0.0004 & 3.2377 & 0.3556 & 0.3582 &
5.769 & 3.481 \\
8 & 0.8271 & 3.2724 & 0.3780 & 0.3807 & 6.120 & 3.646 & & 0.0005 & 3.2715 & 0.3779 & 0.3807 &
6.127 & 3.650 \\
9 & 0.7688 & 3.2582 & 0.3135 & 0.3160 & 6.195 & 3.682 & & 0.0003 & 3.2576 & 0.3135 & 0.3160 &
6.200 & 3.684
\enddata

\tablecomments{Table \ref{dynml} is published in its entirety in the electronic edition of the {\it
Astrophysical Journal}.  A portion is shown here for guidance regarding its form and content.}

\end{deluxetable}

Like Tables~\ref{intmod} and~\ref{glbmod}, Table~\ref{dynml} is arranged in segments, by stellar
mass, $M_i$.  Successive columns list:
\newcounter{tbl3}
\begin{list}{(\arabic{tbl3})}{\usecounter{tbl3}}
\item $k$ --- mass-loss sequence number;
\end{list}
for models with standard mixing-length convective envelopes:
\begin{list}{(\arabic{tbl3})}{\usecounter{tbl3}}
\setcounter{tbl3}{1}
\item $\log R_i$ --- initial radius ($R_\sun$);
\item $M_{\rm KH}$ --- mass threshold at which $\dot{M} = - M/\tau_{\rm KH}$;
\item $\log R_{\rm KH}$ --- Roche lobe radius at which $\dot{M} = - M/\tau_{\rm KH}$;
\item $\log R_{\rm KH}^*$ --- stellar radius when $\dot{M} = - M/\tau_{\rm KH}$;
\item $\zeta_{\rm ad}$ --- critical mass-radius exponent for dynamical time scale mass transfer;
\item $q_{\rm ad}$ --- critical mass ratio for dynamical time scale (conservative) mass transfer;
\end{list}
and for models with artificially isentropic convective envelopes:
\begin{list}{(\arabic{tbl3})}{\usecounter{tbl3}}
\setcounter{tbl3}{7}
\item $\Delta_{\rm exp} \equiv \log (\tilde{R}_i/R_i)$ --- superadiabatic expansion factor;
\item $\tilde{M}_{\rm KH}$ --- mass threshold at which $\dot{M} = - M/\tau_{\rm KH}$;
\item $\log \tilde{R}_{\rm KH}$ --- Roche lobe radius at which $\dot{M} = - M/\tau_{\rm KH}$;
\item $\log \tilde{R}_{\rm KH}^*$ --- stellar radius when $\dot{M} = - M/\tau_{\rm KH}$;
\item $\tilde{\zeta}_{\rm ad}$ --- critical mass-radius exponent for dynamical time scale mass transfer;
and
\item $\tilde{q}_{\rm ad}$ --- critical mass ratio for dynamical time scale (conservative) mass transfer
\end{list}

Columns (3)-(5) and (9)-(11) refer to the points in the critical mass loss sequences at which $\dot{M}$
just reaches $\dot{M}_{\rm KH}$, the characteristic mass loss rate which we identify with the transition
from thermal to dynamical time scale mass transfer. The corresponding initial conditions leading to these
critical points are found in columns (6) and (7) ($\zeta_{\rm ad}$ and $q_{\rm ad}$, respectively) for the
mixing length convection models, and in columns (12) and (13) ($\tilde{\zeta}_{\rm ad}$ and
$\tilde{q}_{\rm ad}$, respectively) for the isentropic convection models.  For reasons outlined above, we
consider the critical mass-radius exponents and mass ratios for these models, $\tilde{\zeta}_{\rm ad}$
and $\tilde{q}_{\rm ad}$ respectively the more realistic, and adopt them in preference to $\zeta_{\rm
ad}$ and $q_{\rm ad}$ below in applying our results to real systems.

It should be noted here that the initial radii listed in Table~\ref{dynml} for the mass-loss sequences differ
slightly from those of the corresponding evolutionary models listed in Table~\ref{glbmod}. By
modifying the surface boundary condition imposed on the adiabatic sequences (see Paper I), we insure
that the luminosity of our mass-losing stars is continuous through the photosphere, and satisfies the
blackbody relation, but at the cost of introducing very small differences in the stellar radii. Comparing
the entries of column (3) of Table~\ref{dynml} with those of column (3) of Table~\ref{glbmod}, the
reader can verify that the difference in $\log R_i$ is in all cases negligible in magnitude, never exceeding
0.0012, and averaging only 0.00004.

\begin{figure}
\plotone{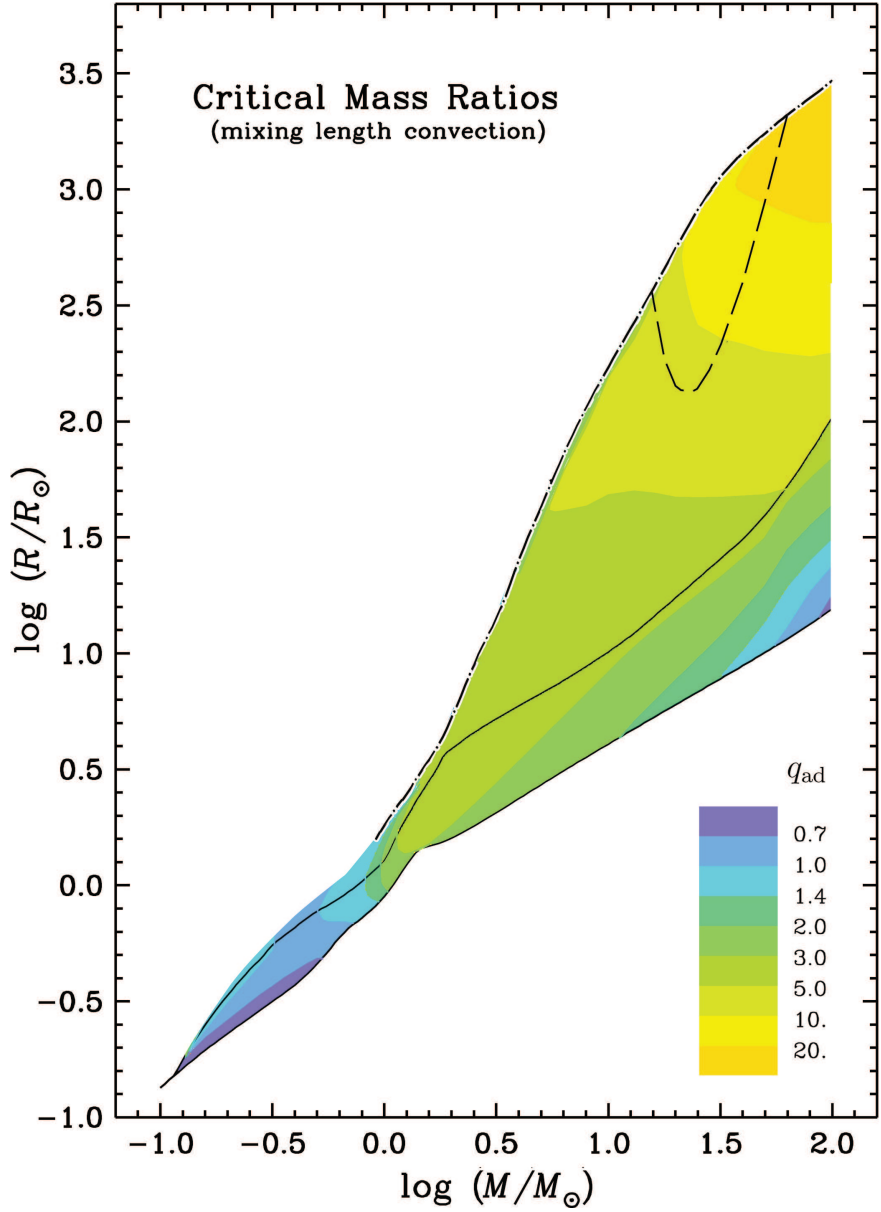}
\caption{Critical mass ratios, $q_{\rm ad}$, for the onset of dynamical time scale mass transfer as
derived from standard evolutionary models, in the mass-radius diagram.\label{qcrit_real}}
\end{figure}

\begin{figure}
\plotone{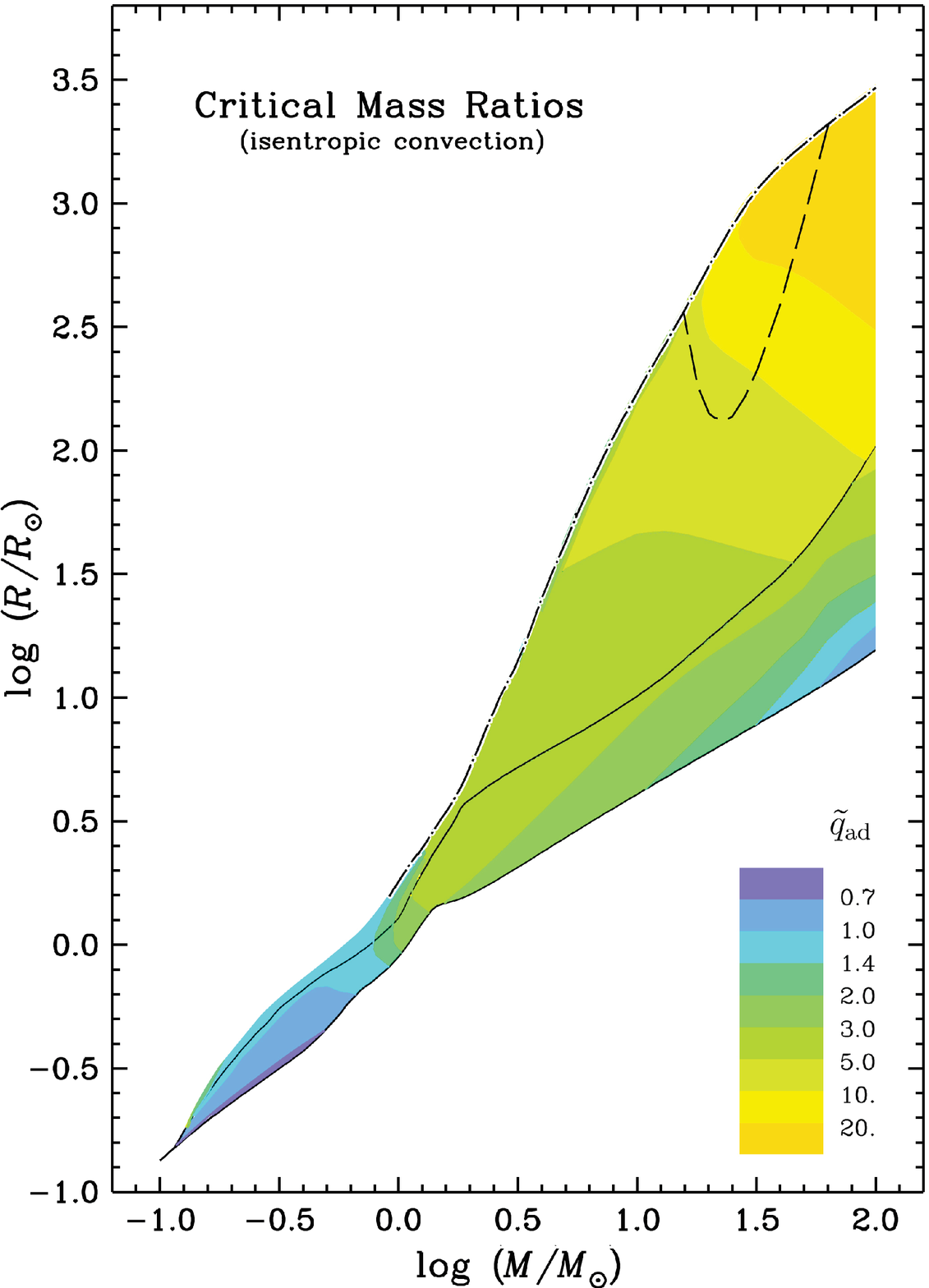}
\caption{Critical mass ratios, $\tilde{q}_{\rm ad}$, for the onset of dynamical time scale mass transfer
as derived from modified evolutionary models with isentropic surface convection zones, in the
mass-radius diagram.  These models mimic the effects of rapid thermal relaxation in the outer layers of
convective stellar envelopes by suppressing the destabilizing effect superadiabatic expansion, thus
providing more realistic estimates of critical mass ratios than the models shown in
Fig.~\ref{qcrit_real}.\label{qcrit_isen}}
\end{figure}

The critical mass ratios found in Table~\ref{dynml} are presented graphically in the form of contour
plots in Figures~\ref{qcrit_real} and~\ref{qcrit_isen} for mixing-length and isentropic envelope models,
respectively. It is immediately apparent that the solutions for $q_{\rm ad}$ and $\tilde{q}_{\rm ad}$
differ very little from each other qualitatively, although $\tilde{q}_{\rm ad}$ is systematically larger
than $q_{\rm ad}$. The difference quantitatively is small except for low-mass main sequence stars,
which have deep but efficient surface convection zones, and for massive stars, where the growing
dominance of radiation pressure throughout their interiors makes their radii very sensitive to small
differences in photospheric density (and entropy).

The most striking feature of Figures~\ref{qcrit_real} and~\ref{qcrit_isen} is the nearly uniform trend
toward larger critical mass ratios with larger radii in the Hertzsprung gap, a feature as well of stars within
the main sequence band itself. These intermediate-mass and massive stars have very thin surface
convection zones, if any at all, and typically contract very rapidly in response to adiabatic mass loss.
Their critical mass ratios for dynamical time scale mass transfer are then set by the \emph{delayed
dynamical instability} described in Paper I, wherein a protracted episode of thermal time scale mass
transfer develops into dynamical instability as mass loss encroaches on the nearly isentropic core of the
donor star. That core, as defined in Table 1, scarcely grows in mass as the star evolves through core
hydrogen burning and contracts toward helium ignition. The growing stellar radius then demands ever
more radical contraction during the thermal mass transfer phase before triggering dynamical instability.

A contributing factor to the increase in $\tilde{q}_{\rm ad}$ with increasing radius is the convergence of
dynamical and thermal time scales for stars of high luminosity with extended envelopes. The ratio of
global stellar thermal to dynamical time scales, $\tau_{\rm th}/\tau_{\rm dyn} \approx (G^3 M^5/R^5
L^2)^{1/2}$, varies from $10^{14}$ in the lower left corner of the mass-radius diagram, to $10^1$ at the
extreme upper right corner of this diagram. The very short thermal time scales for these luminous,
extended stars means that the adiabatic limit can only be reached when the mass transfer rate is extremely
large, and so the relative depth of Roche lobe overflow, $(\tilde{R}_{\rm KH}^* - \tilde{R}_{\rm
KH})/\tilde{R}_{\rm KH}$, needed to reach dynamical mass transfer becomes so large that our 1D
treatment of mass transfer becomes increasingly inadequate.  Indeed, among very luminous stars, the
donors will have overfilled an outer critical surface before reaching the transition to dynamical time scale
mass transfer.  We shall explore this phenomenon, and its implications for the mass transfer process, in a
future study.

\begin{figure}
\epsscale{1.00}
\plotone{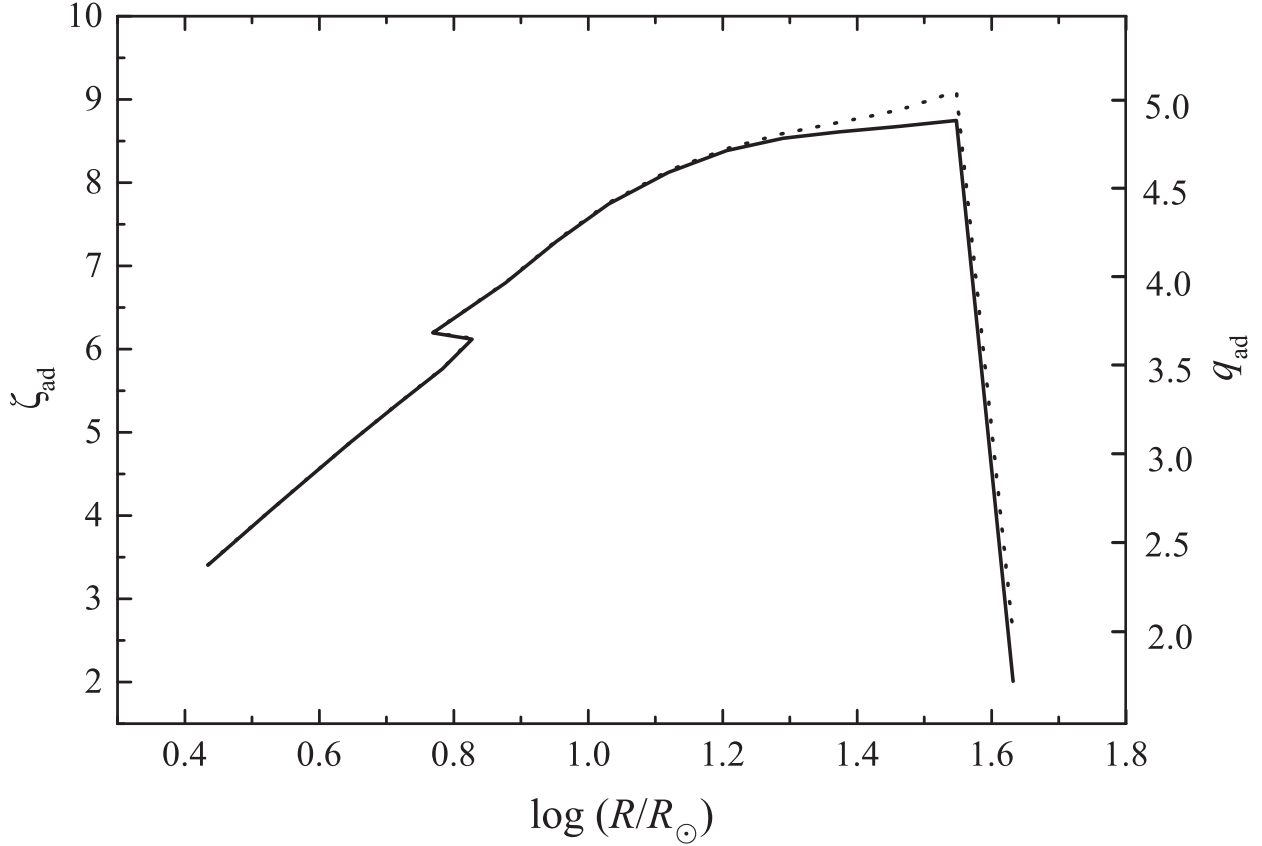}
\caption{The critical mass-radius exponent, $\zeta_{\rm ad}$ and mass ratio $q_{\rm ad}$ as functions
of stellar radius for the $5\ M_\sun$ models shown in Figs.~\ref{5M_Mid_HG} and~\ref{5Msun_Rt},
illustrating the abrupt transition from delayed dynamical instability to prompt dynamical instability at $R
= 35\ R_\sun$, as the star approaches the base of the giant branch.  The solid curve corresponds to
models
with standard mixing-length envelopes ($\zeta_{\rm ad}$ and $q_{\rm ad}$), and the dotted curve to
models with artificially isentropic convective envelopes $\tilde{\zeta}_{\rm ad}$ and $\tilde{q}_{\rm
ad}$.\label{delay_prompt}}
\end{figure}

As intermediate-mass and massive stars approach the base of the giant branch, they develop surface
convection zones that grow rapidly in extent. When the mass of the convective envelope ($M_{\rm ce}$)
reaches approximately $10^{-3} M_i$, the critical conditions for dynamical time scale mass transfer
undergo an abrupt, but continuous, transition from delayed to prompt instability
(Fig.~\ref{delay_prompt}). No longer is dynamical instability preceded by an extended (and extensive)
episode of thermal time scale mass transfer, but mass transfer accelerates directly to dynamical
instability. In Table~\ref{dynml}, cases where critical conditions are set by a prompt
instability can be identified by their small differences between $M_i$ and $M_{\rm KH}$ (or
$\tilde{M}_{\rm KH}$). This difference is much larger for delayed dynamical instability. For the most
luminous, extended stars, however, the convergence of thermal and dynamical time scales greatly blurs
the distinction between prompt and delayed instability.

Main sequence stars with masses $\lesssim 1.1\ {\rm M}_{\sun}$ have surface convection zones of
sufficient depth to be subject to prompt dynamical instability. These surface convection zones increase
rapidly in depth with decreasing main sequence mass, with $\tilde{\zeta}_{\rm ad}$ converging toward
the classical limit for fully convective $n = 3/2$ polytropes ($\zeta_{\rm ad} = - 1/3$, corresponding to
$q_{\rm ad} \approx 2/3$ for main sequence stars of mass ${M \lesssim 0.4\ {\rm
M}_{\sun}}$\footnote{The reader may note that $\zeta_{\rm ad}$ and $q_{\rm ad}$ diverge from the
polytropic limit over this same mass range, and indeed $q_{\rm ad}$ may even become negative. In this
case, all mass ratios would be unstable, and the solutions for $q_{\rm ad}$ are purely formal ones.  For
these stars, thermal time scale mass transfer is so slow that even very modest superadiabatic expansion
can drive mass transfer rates beyond the thermal rate.}.

\section{Comparison with Time-Dependent Mass-Loss Models}
\label{time-dependent}

How well do the threshold mass ratios for dynamical time scale mass transfer, as deduced from the
adiabatic mass loss sequences presented here, replicate the results of time-dependent mass loss
calculations?  To the extent that they concern donor stars within the main sequence or the Hertzsprung
gap, the threshold mass ratios for dynamical time scale mass transfer in common use in binary population
synthesis models \citep[e.g.][]{PZV96,Bel08} derive largely from the early adiabatic mass-loss studies
by \citet{Hje89b}.  Where our models overlap his in mass and evolutionary state, our results are broadly
consistent with his, but of much broader scope.  However, time-dependent calculations suitable for
comparison with our adiabatic mass-loss models are a rarer commodity.  \citet{Iva04} surveyed a
relatively narrow range of parameters ($1 < M_1/M_{\odot} < 3.5$, with $P_{\rm orb} = 1^{\rm d}$ or
$2^{\rm d}$), and within that range, deduced threshold mass ratios in good accord with those presented
here.

A broader, more suitable comparison between adiabatic and time-dependent models is afforded by the
studies of mass transfer from donor stars in the Hertzsprung gap by \citet{Han00} and
\citet{Che02,Che03}.  They modeled mass transfer from donor stars in the mass range $0.0 \le \log
(M_1/M_{\odot}) \le 0.9$, first filling their Roche lobes early, midway, and late in crossing the
Hertzsprung gap.  For each combination of initial mass and radius, time-dependent mass-loss models
were calculated for each of five initial mass ratios ($q_i = 1.1, \;1.5, \;2.0, \;3.0, \;4.0$).  This grid of
models was calculated under various assumptions (with or without convective overshooting and/or mass
and angular momentum loss), of which the set of models with convective overshooting but conservative
mass transfer \citet{Che03} correspond most closely to the assumptions adopted in our adiabatic mass
loss models.  The underlying stellar structure code employed by \citeauthor{Che03} shares the same
basic platform as that used to generate our family of initial models, differing significantly only in the
algorithm used to calculate mass transfer rates from the degree to which the donor star overfills its Roche
lobe.  The \citeauthor{Che03} survey does not specifically aim to quantify critical mass ratios for
dynamical time scale mass transfer, but it does identify cases in which the initial models succumb
directly to dynamical mass transfer (prompt dynamical instability), those which become unstable as they
reach the red giant branch (delayed dynamical instability), and those which remain in stable mass transfer
-- thermal or nuclear time scale -- throughout.  These results can be used to bracket the critical mass ratio
for dynamical time scale mass transfer, for comparison with our results, as shown in
Table~\ref{comparison}.

\begin{deluxetable}{cccccccc}
\tabletypesize{\scriptsize}
\tablecolumns{8}
\tablewidth{0pt}
\tablecaption{Comparison of $\tilde{q}_{\rm ad}$ with time-dependent models
\citep{Che03}\label{comparison}}
\tablehead{
\colhead{$M/M_{\odot}$} & \colhead{$\log L/L_{\odot}$} & \colhead{$\log R/R_{\odot}$} &
\colhead{$\log T_e/{\rm K}$} & \colhead{$q_\ell$} & \colhead{$\tilde{q}_{\rm ad}$} &
\colhead{$q_u$} & \colhead{mode}}
\startdata
1.000 & 0.240 & 0.140 & 3.752 & 2.000 & 1.984 & 3.000 & P \\
1.000 & 0.332 & 0.218 & 3.736 & 1.500 & 1.452 & 2.000 & P \\
1.000 & 0.354 & 0.298 & 3.701 & 1.100 & 0.968 & 1.500 & P \\
1.259 & 0.688 & 0.322 & 3.773 & 2.000 & 3.300 & 3.000 & D \\
1.259 & 0.691 & 0.357 & 3.756 & 2.000 & 2.941 & 3.000 & D \\
1.259 & 0.609 & 0.392 & 3.718 & 1.100 & 1.073 & 1.500 & P \\
1.585 & 1.088 & 0.492 & 3.788 & 3.000 & 3.825 & 4.000 & D \\
1.585 & 1.107 & 0.522 & 3.778 & 3.000 & 3.406 & 4.000 & D \\
1.585 & 1.095 & 0.556 & 3.758 & 2.000 & 2.570 & 3.000 & D \\
1.995 & 1.531 & 0.654 & 3.818 & 4.000 & 4.570 & & T \\
1.995 & 1.545 & 0.701 & 3.797 & 4.000 & 4.616 & & T \\
1.995 & 1.538 & 0.746 & 3.773 & 3.000 & 4.216 & 4.000 & D \\
2.512 & 1.907 & 0.724 & 3.877 & 3.000 & 4.545 & 4.000 & D \\
2.512 & 1.982 & 0.835 & 3.840 & 4.000 & 4.738 & & T \\
2.512 & 1.956 & 0.945 & 3.778 & 4.000 & 4.713 & & T \\
3.162 & 2.381 & 0.800 & 3.957 & 4.000 & 4.344 & & T \\
3.162 & 2.403 & 0.977 & 3.874 & 4.000 & 4.625 & & T \\
3.162 & 2.345 & 1.155 & 3.770 & 4.000 & 4.482 & & T \\
3.981 & 2.791 & 0.865 & 4.027 & 3.000 & 4.281 & 4.000 & D \\
3.981 & 2.812 & 1.121 & 3.904 & 4.000 & 4.758 & & T \\
3.981 & 2.730 & 1.374 & 3.757 & 4.000 & 4.170 & & T \\
5.012 & 3.132 & 0.903 & 4.093 & 3.000 & 4.138 & 4.000 & D \\
5.012 & 3.205 & 1.244 & 3.941 & 4.000 & 4.748 & & T \\
5.012 & 3.109 & 1.589 & 3.744 & 4.000 & 4.793 & & T \\
6.310 & 3.536 & 0.974 & 4.159 & 3.000 & 3.980 & 4.000 & D \\
6.310 & 3.577 & 1.385 & 3.963 & 4.000 & 4.813 & & T \\
6.310 & 3.482 & 1.796 & 3.734 & 4.000 & 6.135 & & T \\
7.943 & 3.887 & 1.045 & 4.211 & 2.000 & 3.875 & & D \\
7.943 & 3.924 & 1.515 & 3.985 & 4.000 & 4.856 & & T \\
7.943 & 3.828 & 1.987 & 3.726 & 3.000 & 5.841 & 4.000 & D \\
\enddata
\end{deluxetable}

Table~\ref{comparison} summarizes the constraints on the threshold mass ratio for dynamical time scale
mass transfer as inferred from \citet{Che03}.  Each line of this table refers to a family of time-dependent
mass loss calculations spanning the five trial mass ratios identified above, but
sharing a common evolutionary state for the donor star.  Columns (1) through (4) list donor mass,
luminosity, radius, and effective temperature (averaged over minor variations in $\log L$, $\log R$, and
$\log T_e$ among the trial mass ratios).  Since the susceptibility to dynamical instability increases with
increasing mass ratio of donor to accretor, the largest mass ratio to \emph{avoid} dynamical instability in
the time-dependent calculations presumably sets a lower limit ($q_\ell$ --- column (5) ) to the threshold
mass ratio for dynamical time scale mass transfer, while the smallest mass ratio to \emph{trigger}
dynamical instability sets an upper limit ($q_u$ --- column (7) to that threshold mass ratio.  Critical mass
ratios interpolated from our adiabatic mass loss sequences ($\tilde{q}_{\rm ad}$) are found in column
(6).  The final column (8) of Table~\ref{comparison} identifies the nature of the mass transfer instability:
$P$ --- prompt dynamical instability; $D$ --- delayed dynamical instability; and $T$ --- thermal time
scale instability.

The close agreement demonstrated in Table~\ref{comparison} between time-dependent and adiabatic
thresholds for dynamical instability give confidence that the approximations inherent in the adiabatic
approach are of mior consequence.  While our results do not always satisfy the expected inequality,
$q_\ell < \tilde{q}_{\rm ad} < q_u$, $\tilde{q}_{\rm ad}$ rarely strays as much as 10\% outside those
bounds, except for the very most massive and luminous stars included in the \citeauthor{Che03} survey.
The singular exception to this close agreement between adiabatic and time-dependent critical mass ratios
occurs for the very most luminous and extended model in the \citeauthor{Che03} survey.  We attribute
this discrepancy to a shortcoming in the prescription used in their studies to relate $\dot{M}$ to the
extent of Roche lobe overflow.  While adequate when that overflow extent $(R - R_{\rm L})/R_{\rm
L})$ is small, their prescription for $\dot{M}$ breaks down for stars with extended envelopes, as it fails
to reflect the natural dynamical time scale, $\tau_{\rm dyn} \sim (G\rho)^{-1/2}  \propto P_{\rm orb}$
\citep[cf.][pp.~132-134]{Egg06}.  It thus overestimates mass transfer rates and the propensity toward
dynamical instability in that limit.  As we shall demonstrate in the next installment in the present series
of papers, a physically realistic model for $\dot{M}$ is essential in evaluating critical mass ratios for
stars with extended envelopes.

\section{An Application: Cataclysmic Variable Stars}
\label{appl}

\begin{deluxetable}{lrclllcc}
\tabletypesize{\scriptsize}
\tablewidth{0pt}
\tablecolumns{8}
\tablecaption{CVs with robust WD mass determinations\label{CVmasses}}
\tablehead{
\colhead{System} & \colhead{$P_{\rm orb}$} & \colhead{Source\tablenotemark{a}} & \colhead
{$M_{\rm wd}$} & \colhead{$M_2$} & \colhead{$q$} & \colhead{Method\tablenotemark{b}} &
\colhead{Refs} \\
\colhead{} & \colhead{(min)} & \colhead{} & \colhead{($M_\sun$)} & \colhead{($M_\sun$)} &
\colhead{} & \colhead{} & \colhead{}}
\startdata
\object{OV Boo}\tablenotemark{c} & \phn\phn66.6 & ZSG & 0.892$\:\pm\:$0.008 &
\phm{>}0.0575$\:\pm\:$0.0020 & \phm{>}0.0647$\:\pm\:$0.0018 & e & 1,2,3 \\
\object{SDSS1433+1011} & 78.1 & ZSG & 0.865$\:\pm\:$0.005 & \phm{>}0.0571$\:\pm\:$0.0007 &
\phm{>}0.1115$\:\pm\:$0.0016 & e & 1,4,5 \\
\object{WZ Sge} & 81.6 & ZSG & 0.85\phn$\:\pm\:$0.04 & \phm{>}0.078\phn$\:\pm\:$0.006 &
\phm{>}0.092\phn$\:\pm\:$0.008 & d,g,sp & 6,7 \\
\object{SDSS1501+5501} & 81.9 & ZSG & 0.767$\:\pm\:$0.027 & \phm{>}0.077\phn$\:\pm\:$0.010 &
\phm{>}0.101\phn$\:\pm\:$0.010 & e & 1,4 \\
\object{SDSS1035+0551} & 82.1 & ZSG & 0.350$\:\pm\:$0.009 & \phm{>}0.0475$\:\pm\:$0.0012 &
\phm{>}0.0571$\:\pm\:$0.0010 & e & 1,8 \\
\object{NZ Boo}\tablenotemark{d} & 84.8 & ZSG & 0.709$\:\pm\:$0.004 &
\phm{>}0.0781$\:\pm\:$0.0008 & \phm{>}0.1099$\:\pm\:$0.0007 & e & 1,4 \\
\object{SDSS0903+3300} & 85.1 & ZSG & 0.872$\:\pm\:$0.011 & \phm{>}0.099\phn$\:\pm\:$0.004 &
\phm{>}0.113\phn$\:\pm\:$0.004 & e & 1,4 \\
\object{XZ Eri} & 88.1 & ZSG & 0.769$\:\pm\:$0.017 & \phm{>}0.091\phn$\:\pm\:$0.004 &
\phm{>}0.118\phn$\:\pm\:$0.003 & e & 1,9 \\
\object{SDSS1227+5139} & 90.7 & ZSG & 0.796$\:\pm\:$0.018 & \phm{>}0.0889$\:\pm\:$0.0025 &
\phm{>}0.1115$\:\pm\:$0.0016 & e & 1 \\
\object{OY Car} & 90.9 & ZSG & 0.840$\:\pm\:$0.040 & \phm{>}0.086\phn$\:\pm\:$0.005 &
\phm{>}0.102\phn$\:\pm\:$0.003 & e & 4,10 \\
\object{DI Phe}\tablenotemark{e} & 94.4 & ZSG & 0.935$\:\pm\:$0.031 &
\phm{>}0.101\phn$\:\pm\:$0.003 & \phm{>}0.1097$\:\pm\:$0.0008 & e & 1 \\
\object{SDSS1152+4049} & 97.5 & ZSG & 0.560$\:\pm\:$0.028 & \phm{>}0.087\phn$\:\pm\:$0.006 &
\phm{>}0.155\phn$\:\pm\:$0.006 & e & 1 \\
\object{EX Hya} & 98.3 & WR & 0.484$\:\pm\:$0.393 & \phm{>}0.080\phn$\:\pm\:$0.054 &
\phm{>}0.166\phn$\:\pm\:$0.075 & d,e & \phm{00,00,}36,37,38\phm{,00,00} \\
\object{OU Vir} & 104.7 & ZSG & 0.703$\:\pm\:$0.012 & \phm{>}0.1157$\:\pm\:$0.0022 &
\phm{>}0.1641$\:\pm\:$0.0013 & e & 1,11,12 \\
\object{HT Cas} & 106.1 & ZSG & 0.61\phn$\:\pm\:$0.04 & \phm{>}0.09\phn\phn$\:\pm\:$0.02 &
\phm{>}0.15\phn\phn$\:\pm\:$0.03 & e & 13 \\
\object{HT Cas} & 106.1 & WR & 0.842$\:\pm\:$0.099 & \phm{>}0.124\phn$\:\pm\:$0.032 &
\phm{>}0.147\phn$\:\pm\:$0.032 & d,r,e & 39,40,13 \\
\object{IY UMa} & 106.4 & ZSG & 0.79\phn$\:\pm\:$0.04 & \phm{>}0.10\phn\phn$\:\pm\:$0.01 &
\phm{>}0.125\phn$\:\pm\:$0.008 & e & 14 \\
\object{VW Hyi} & 107.0 & ZSG & 0.71\phn$\:\pm\:$0.22 & \phm{>}0.11\phn\phn$\:\pm\:$0.03 &
\phm{>}0.148\phn$\:\pm\:$0.004 & g & 15,16 \\
\object{Z Cha} & 107.3 & ZSG & 0.84\phn$\:\pm\:$0.09 & \phm{>}0.125\phn$\:\pm\:$0.014 &
\phm{>}0.20\phn\phn$\:\pm\:$0.02 & e,d & 17,18 \\
\object{Z Cha} & 107.3 & WR & 0.857$\:\pm\:$0.181 & \phm{>}0.122\phn$\:\pm\:$0.026 &
\phm{>}0.142\phn$\:\pm\:$0.003 & d,e & 41,17,18 \\
\object{DV UMa} & 123.6 & ZSG & 1.098$\:\pm\:$0.024 & \phm{>}0.196\phn$\:\pm\:$0.005 &
\phm{>}0.1778$\:\pm\:$0.0022 & e & 1,9 \\
\object{V1258 Cen}\tablenotemark{f} & 128.1 & ZSG & 0.736$\:\pm\:$0.014 &
\phm{>}0.177\phn$\:\pm\:$0.021 & \phm{>}0.240\phn$\:\pm\:$0.021 & e & 1 \\
\object{V1239 Her}\tablenotemark{g} & 144.1 & ZSG & 0.91\phn$\:\pm\:$0.03 &
\phm{>}0.223\phn$\:\pm\:$0.010 & \phm{>}0.248\phn$\:\pm\:$0.005 & e & 1,19 \\
\object{AM Her} & 185.7 & ZSG & 0.78\phn$\:\pm\:$0.15 &
\phm{>0.223}\nodata & \phm{>0.248}\nodata & sp & 20 \\
% \phm{>}0.18\phn\phn$\:\pm\:$0.01 & \phm{>}0.23\phn\phn$\:\pm\:$0.05 & sp & 20 \\
\object{DW UMa} & 196.7 & ZSG & 0.87\phn$\:\pm\:$0.19 & $\!\!>$0.16 & $\!\!>$0.24 & e & 21 \\
\object{IP Peg} & 227.8 & ZSG & 1.16\phn$\:\pm\:$0.02 & \phm{>}0.55\phn\phn$\:\pm\:$0.02 &
\phm{>}0.48\phn\phn$\:\pm\:$0.01 & e & 22 \\
\object{IP Peg} & 227.8 & WR & 1.032$\:\pm\:$0.100 & \phm{>}0.416\phn$\:\pm\:$0.042 &
\phm{>}0.403\phn$\:\pm\:$0.014 & d,r,e & 42,43,44,45,40,46,47 \\
\object{GY Cnc} & 252.6 & ZSG & 0.99\phn$\:\pm\:$0.12 & \phm{>}0.38\phn\phn$\:\pm\:$0.06 &
\phm{>}0.387\phn$\:\pm\:$0.031 & e & 23 \\
\object{GY Cnc} & 252.6 & WR & 0.892$\:\pm\:$0.146 & \phm{>}0.366\phn$\:\pm\:$0.071 &
\phm{>}0.410\phn$\:\pm\:$0.050 & d,e & 48,23 \\
\object{U Gem} & 254.7 & ZSG & 1.20\phn$\:\pm\:$0.09 & \phm{>}0.42\phn\phn$\:\pm\:$0.04 &
\phm{>}0.35\phn\phn$\:\pm\:$0.05 & d,g,sp & 24,25,26,27,28,29 \\
\object{U Gem} & 254.7 & WR & 0.982$\:\pm\:$0.255 & \phm{>}0.352\phn$\:\pm\:$0.057 &
\phm{>}0.359\phn$\:\pm\:$0.041 & d,e & 49,27,25 \\
\object{BD Pav} & 258.2 & WR & 0.962$\:\pm\:$0.100 & \phm{>}0.466\phn$\:\pm\:$0.100 &
\phm{>}0.485\phn$\:\pm\:$0.064 & d,r,e & 50,49 \\
\object{SDSS1006+2337} & 267.7 & ZSG & 0.78\phn$\:\pm\:$0.12 &
\phm{>}0.40\phn\phn$\:\pm\:$0.10 & \phm{>}0.51\phn\phn$\:\pm\:$0.08 & e & 30 \\
\object{DQ Her} & 278.8 & ZSG & 0.60\phn$\:\pm\:$0.07 & \phm{>}0.40\phn\phn$\:\pm\:$0.05 &
\phm{>}0.66\phn\phn$\:\pm\:$0.04 & d & 31 \\
\object{DQ Her} & 278.8 & WR & 0.593$\:\pm\:$0.128 & \phm{>}0.369\phn$\:\pm\:$0.082 &
\phm{>}0.623\phn$\:\pm\:$0.099 & d,r,e & 31,51 \\
\object{EX Dra} & 302.3 & WR & 0.696$\:\pm\:$0.120 & \phm{>}0.464\phn$\:\pm\:$0.097 &
\phm{>}0.666\phn$\:\pm\:$0.076 & d,r,e & 52,53,54 \\
\object{RW Tri} & 333.9 & WR & 0.618$\:\pm\:$0.219 & \phm{>}0.456\phn$\:\pm\:$0.152 &
\phm{>}0.739\phn$\:\pm\:$0.116 & d,r,e & 55,56,57,58 \\
\object{V347 Pup} & 334.0 & ZSG & 0.63\phn$\:\pm\:$0.04 & \phm{>}0.52\phn\phn$\:\pm\:$0.06 &
\phm{>}0.83\phn\phn$\:\pm\:$0.05 & d & 32 \\
\object{V347 Pup} & 334.0 & WR & 0.616$\:\pm\:$0.041 & \phm{>}0.497\phn$\:\pm\:$0.050 &
\phm{>}0.806\phn$\:\pm\:$0.049 & d,r,e & 59,32 \\
\object{EM Cyg} & 418.9 & ZSG & 1.00\phn$\:\pm\:$0.06 & \phm{>}0.77\phn\phn$\:\pm\:$0.08 &
\phm{>}0.77\phn\phn$\:\pm\:$0.04 & d & 33 \\
\object{EM Cyg} & 418.9 & WR & 1.026$\:\pm\:$0.069 & \phm{>}0.903\phn$\:\pm\:$0.099 &
\phm{>}0.880\phn$\:\pm\:$0.052 & d,r,e & 60,61,62 \\
\object{AC Cnc} & 432.7 & ZSG & 0.76\phn$\:\pm\:$0.03 & \phm{>}0.77\phn\phn$\:\pm\:$0.05 &
\phm{>}1.02\phn\phn$\:\pm\:$0.04 & d & 34 \\
\object{AC Cnc} & 432.7 & WR & 0.760$\:\pm\:$0.042 & \phm{>}0.774\phn$\:\pm\:$0.044 &
\phm{>}1.018\phn$\:\pm\:$0.052 & d,r,e & 63,34 \\
\object{V363 Aur} & 462.6 & ZSG & 0.90\phn$\:\pm\:$0.06 & \phm{>}1.06\phn\phn$\:\pm\:$0.11 &
\phm{>}1.17\phn\phn$\:\pm\:$0.07 & d & 34 \\
\object{V363 Aur} & 462.6 & WR & 0.898$\:\pm\:$0.094 & \phm{>}1.039\phn$\:\pm\:$0.097 &
\phm{>}1.157\phn$\:\pm\:$0.108 & d,r,e & 64,34 \\
\object{BT Mon} & 480.7 & WR & 1.062$\:\pm\:$0.218 & \phm{>}0.914\phn$\:\pm\:$0.086 &
\phm{>}0.861\phn$\:\pm\:$0.164 & d,r,e & 65,66 \\
\object{AE Aqr} & 592.8 & ZSG & 0.63\phn$\:\pm\:$0.05 & \phm{>}0.37\phn\phn$\:\pm\:$0.04 &
\phm{>}0.60\phn\phn$\:\pm\:$0.02 & d & 35 \\
\object{AE Aqr} & 592.8 & WR & 0.864$\:\pm\:$0.035 & \phm{>}0.609\phn$\:\pm\:$0.054 &
\phm{>}0.704\phn$\:\pm\:$0.034 & d,r & \phm{00,00,}67,68,69\phm{,00,00} \\
\object{U Sco} & 1772.0 & WR & 1.501$\:\pm\:$0.485 & \phm{>}0.821\phn$\:\pm\:$0.231\phn &
\phm{>}0.547\phn$\:\pm\:$0.102\phn & d,e & 70,71 \\
\enddata

\tablenotetext{a}{ZSG --- compilation by \citet{Zor11}; WR --- analysis by \citet{Web05}}
\tablenotetext{b}{Basis of analysis: (e) --- eclipse light curves; (d) --- radial velocity curves; (g) ---
gravitational redshifts; (sp) --- spectrophotometric modeling; (r) --- rotational velocity of donor star}
\tablenotetext{c}{SDSS1507+5230}
\tablenotetext{d}{SDSS1502+3334}
\tablenotetext{e}{CTCV2354--4700}
\tablenotetext{f}{CTCV1300--3052}
\tablenotetext{g}{SDSS1702+3229}

\tablerefs{(1)\citealt{Sav11}; (2) \citealt{Lit07}; (3) \citealt{Pat08}; (4) \citealt{Lit08}; (5)
\citealt{Tul09}; (6) \citealt{Ste07}; (7) \citealt{Lon04}; (8) \citealt{Lit06b}; (9) \citealt{Fel04c}; (10)
\citealt{Woo89}; (11) \citealt{Fel04a}; (12) \citealt{Fel04b}; (13) \citealt{Hor91}; (14) \citealt{Ste03};
(15) \citealt{Smi06}; (16) \citealt{Sio97}; (17) \citealt{Wad88}; (18) \citealt{WHB86}; (19)
\citealt{Lit06a}; (20) \citealt{Gan06}; (21) \citealt{Ara03}; (22) \citealt{Cop10}; (23) \citealt{Tho00};
(24) \citealt{Ech07}; (25) \citealt{Zha87}; (26) \citealt{Sio98}; (27) \citealt{Lon99}; (28)
\citealt{Nay05}; (29) \citealt{Lon06}; (30) \citealt{Sou09}; (31) \citealt{Hor93}; (32) \citealt{Tho05};
(33) \citealt{Wel07}; (34) \citealt{Tho04}; (35) \citealt{Ech08}; (36) \citealt{Bel03}; (37)
\citealt{VdP03}; (38) \citealt{Muk98}; (39) \citealt{You81}; (40) \citealt{Cat99}; (41) \citealt{Mar87};
(42) \citealt{Hes89}; (43) \citealt{Sma02}; (44) \citealt{Bee00}; (45) \citealt{Mar89}; (46)
\citealt{Mar88}; (47) \citealt{Woo86}; (48) \citealt{Sha00}; (49) \citealt{Fri90}; (50) \citealt{Axe88};
(51) \citealt{Sma80}; (52) \citealt{Fie94}; (53) \citealt{Bil96}; (54) \citealt{Bap00}; (55)
\citealt{Sti95}; (56) \citealt{Poo03}; (57) \citealt{Hor85}; (58) \citealt{Sma95}; (59) \citealt{Sti98};
(60) \citealt{Sto81}; (61) \citealt{Nor00}; (62) \citealt{Mum69}; (63) \citealt{Sch84}; (64)
\citealt{Sch86}; (65) \citealt{Smi98}; (66) \citealt{Rob82}; (67) \citealt{deJ94}; (68) \citealt{Era94};
(69) \citealt{Cas96}; (70) \citealt{Sch95}; (71) \citealt{Tho01}}

\end{deluxetable}

\begin{figure}
\includegraphics[angle=270,scale=.70]{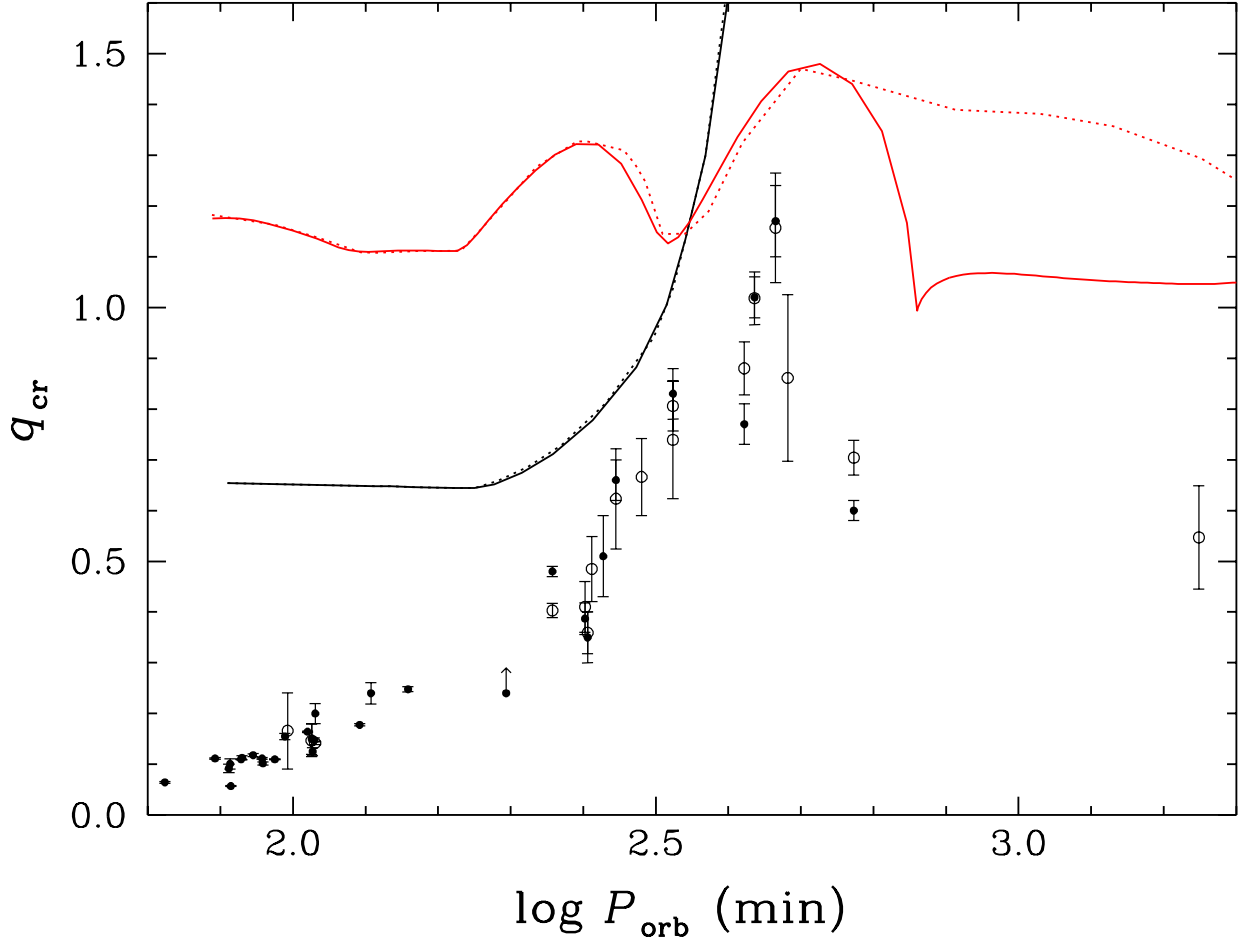}
\caption{The mass ratio distribution for cataclysmic variables with robust mass determinations.  Filled
circles denote masses drawn from \citet{Zor11}, and open circles masses drawn from an unpublished
analysis by \citet{Web05}.  Black curves map upper limits to the mass ratio for stability against
dynamical time scale mass transfer; red curves map the corresponding upper limits to the mass ratio
against thermal time scale mass transfer.  Solid lines mark limits for donor zero-age main sequence stars;
dotted lines the upper envelopes of limits for evolved donors.\label{CVs}}
\end{figure}

Our stability limits reproduce with considerable fidelity the observed range of mass ratios as a function
of orbital period for cataclysmic variables with reliable structural parameters. We adopted the sample of
32 CVs from \citet{Zor11} judged by them to have robust mass determinations, supplemented by 17 CVs
(11 in common with Zorotovic et al.) from an unpublished study by \citet{Web05}, as detailed in
Table~\ref{CVmasses}. These systems must be stable against dynamical and thermal time scale mass
transfer, and indeed \emph{all} of them lie within the bounds permitted by the dynamical and thermal
stability limits (Fig.~\ref{CVs}). As expected, those systems with mass ratios $q > 1$ lie in the period
range ${5\fh 8} < P_{\rm orb} < 12^{\rm h}$ where stability constraints are weakest. Among
shorter-period systems ($P_{\rm orb} < 4\fh 5$), the observed mass ratios fall progressively further
below our derived stability limits. (In this period range, stability against dynamical time scale mass
transfer poses the stronger constraint.) We interpret this divergence as an artifact of pre-CV common
envelope evolution, which disfavors survival of systems producing low-mass white dwarfs.

The dynamical and thermal stability limits derived here assume conservation of mass and orbital angular
momentum. The fact that observed systems appear to obey these limits implies that any temporary
accumulation of angular momentum in the accretion disk or in rotation of the accreting white dwarf must
be restored to the binary orbit on a time scale short compared with the growth time scale of any mass
transfer instability triggered by nonconservative mass transfer.

\section{Discussion and Conclusions}
\label{discussion}

This study is the first of a pair (the second to deal with stars on the giant and asymptotic giant branches)
that attempt for the first time to survey systematically the thresholds for dynamical time scale mass
transfer over the entire span of possible donor star evolutionary states. These thresholds mark bifurcation
points in close binary evolution, separating evolutionary channels proceeding on a thermal time scale (or
slower) from those proceeding on a (typically) far more rapid time scale leading to common envelope
evolution. Its most obvious immediate application, then, is as input to population synthesis studies of
close binary evolution that seek to quantify the frequency and properties of various possible evolutionary
channels.

We are confident that the families of adiabatic mass-loss calculations presented here not only capture the
qualitative trends of dynamical thresholds with evolutionary state of the donor, but are quantitatively
reliable so long as the donor's dynamical time scale is much shorter than its thermal time scale (justifying
the adiabatic approximation). Where we can test our results against observational constraints, as
exemplified above in the application to cataclysmic variables, they appear robust, but such direct
comparisons are very rarely possible. Some cautionary remarks concerning the limitations of our
calculations are therefore warranted.

Foremost among the approximations we have employed is the treatment of the donor response to mass
loss as one of adiabatic expansion throughout the donor interior. This approximation may be valid
throughout the bulk of the interior once the mass loss rate significantly surpasses the thermal time scale
rate, but as noted above it must break down near the stellar photosphere, where radiative relaxation
becomes extremely rapid. More significantly, one must recognize that the growth to supra-thermal mass
transfer rates generally extends over many thermal time scales, while we approximate the donor response
even in these circumstances as purely adiabatic. An estimate of the amount of mass lost during the
acceleration to dynamical time scale can be had from the difference $M - M_{\rm KH}$ in
Table~\ref{dynml}. For stars with moderately deep surface convection zones, thermal
relaxation during this acceleration phase is probably of little consequence because convection zones tend
to respond as coherent entities (specific entropy rises or falls more or less uniformly throughout the
convection zone), but more significantly because such stars are subject to prompt dynamical instabilities
that cut short the acceleration phase. Among stars with radiative envelopes, on the other hand, dynamical
instability is generally of the delayed variety, and thermal relaxation during the long run up to dynamical
instability may be extensive.  That relaxation typically involves absorption of a large fraction of the
interior luminosity in the outer envelope, as described above in Section~\ref{response} and illustrated in
Fig.~\ref{5M_Mid_HG}. That energy absorption is directed toward rebuilding the strong, positive
entropy gradient in the outer envelope that characterizes the structure of radiative stars in thermal
equilibrium (see Fig. 1 in Paper I). Since thermal relaxation in radiative envelopes drives expansion of
the surface layers (relative to purely adiabatic expansion), it tends to drive higher mass transfer rates than
we would calculate from our adiabatic models. Those higher rates may tend to drive the donor toward
dynamical instability, but for reasons we elaborate following, we expect that in most cases mass transfer
will be cut short by contact with the accreting star before a delayed dynamical instability is manifested.

It is quite likely that the great majority of binaries with mass ratios exceeding $\tilde{q}_{\rm ad}$ for
the delayed dynamical instability will in fact evolve into contact before actually reaching the point of
instability if they have nondegenerate accretors. Because $\tilde{q}_{\rm ad}$ is in most cases relatively
large ($\tilde{q}_{\rm ad} > 3$), orbital contraction during mass transfer is severe, whereas the accreting
stars tend to expand far beyond their thermal equilibrium radii. In their survey of case A mass transfer
(mass transfer initiated while the donor star is still in central hydrogen burning), \citet{Nel01} found that
practically all intermediate-mass donors with $q > \tilde{q}_{\rm ad}$ suffered this fate. Since
$\tilde{q}_{\rm ad}$ increases as stars evolve from the terminal main sequence to the base of the giant
branch, while orbital contraction during mass transfer scales very roughly as $q^{-1}$, these more
evolved donors likely also reach contact before developing dynamical instability. If the accretor is
compact, it is likely that rapid mass transfer is very non-conservative (super-Eddington winds tend to be
strongly stabilizing); it remains to be seen whether delayed dynamical instabilities can be manifested in
this case.

The most massive, extended stars in this survey have dynamical time scales that are scarcely a factor of
$10^2$ shorter than their thermal time scales. In reality, stars in the upper reaches of our mass range
generally suffer quite extensive mass loss in stellar winds, and tend to intrinsic variability as luminous
blue variables. Those losses and variability are neglected here, and while our results may still be useful in
framing expectations for the behavior of these stars as donors, they are unlikely to be quantitatively
reliable.

Finally, with respect to the dynamical response of a binary orbit to mass transfer and mass loss, we again
emphasize that, for the sake of clarity and economy, we have assumed conservation of total mass and of
total orbital angular momentum, neglecting rotational contributions to the total angular momentum of the
binary, and adopted the usual approximations for the tidal limit (Roche lobe) of the donor star. Of course,
in reality the response of the donor star's Roche lobe to mass transfer depends on systemic losses of mass
and angular momentum, as well as on angular momentum transport within the binary. At this juncture, no
robust theory exists for quantifying those processes, and they are typically parameterized using ad hoc
prescriptions for the fraction of mass lost by the donor, but retained by the accretor, the specific angular
momentum carried away by systemic mass loss, and the coupling between stellar rotation and the binary
orbit. All of these processes introduce additional dimensions to the problem of quantifying thresholds for
dynamical time scale mass transfer.

Given a prescription for how the donor Roche lobe responds to mass loss, our adiabatic mass loss
sequences are in principle applicable to non-conservative mass transfer as well. It bears emphasizing that
the adiabatic mass-radius exponent, $\tilde{\zeta}_{\rm ad}$, is intrinsic to the donor star. Within the
context of the approximations employed in this study, it depends on the binary mass ratio only through
the function $f(q)$ in Paper I (Eqn. A10), which dependence is extremely weak. Limiting mass ratios for
dynamical stability in the case of non-conservative mass transfer, $q_{\rm ad}^{\rm (nc)}$, can therefore
be
calculated by solving the relationship
\[
\zeta_{\rm L}(q_{\rm ad}^{\rm (nc)}) = \tilde{\zeta}_{\rm ad},
\]
where
\[
\zeta_{\rm L}(q) = \left( \frac{\partial \ln R_{\rm L}(q)}{\partial \ln M} \right)_{\rm nc}
\]
is the Roche lobe mass-radius exponent appropriate to the adopted non-conservative treatment of mass
transfer. If the donor is subject to prompt dynamical instability (possesses a non-negligible surface
convection zone), the resulting value of $q_{ad}^{\rm (nc)}$ should be robust, as we can then neglect
higher-order terms in the dependence of $\ln R$ and $\ln R_{\rm L}$ on $\ln M$ in the initial phases of
mass transfer.  If the donor is subject to delayed dynamical instability, on the other hand, the solution for
$q_{\rm ad}^{\rm (nc)}$ is may be subject to larger systematic errors, depending on the details of the
adopted non-conservative treatment.  Greater accuracy then requires detailed knowledge of $R(M)$
along the adiabatic mass-loss sequence. The necessary details are available from the authors upon
request.\footnote{Tables~\ref{intmod}-\ref{dynml} are combined in a single, machine-readable table
available in the online journal.}

In the next (third) installment in this series of papers, we will take up the adiabatic responses of stars
with convective envelopes --- those on the giant and asymptotic giant branches. Those models present
new issues of interpretation, but are prime candidates for systems prone to common envelope evolution.
The following (fourth) installment will deal directly with the energetics of common envelope evolution,
circumscribing conditions under which survival of common envelope evolution is energetically allowed.
It will be followed by a survey of critical conditions for the onset of thermal time scale mass transfer,
some initial results of which were employed in the survey of cataclysmic variable stability discussed
above.

\acknowledgments

HG gratefully acknowledges the kind help and constant encouragement of RW and ZH. This work was
supported by grants from the National Natural Science Foundation of China (NSFC) (Nos. 11203065,
11033008, 11390374), the Natural Science Foundation of Yunnan Province (Grant No. 2014FB189),
West Light Foundation of the Chinese Academy of Sciences (CAS), and the Department of Astronomy,
University of Illinois at Urbana-Champaign. ZH is partly supported by the Science and Technology
Innovation Talent Programme of the Yunnan Province (Grant No. 2013HA005) and the CAS (Grant No.
XDB09010202).  XC is partly supported by NSFC (Grant Nos. 11422324, 11173055), and the Talent
Project of Young Researchers of Yunnan Province (Grant No. 2012HB037).  HG thanks CAS and the
Department of Astronomy, University of Illinois at Urbana-Champaign for the a one year visiting
appointment. RFW thanks Yunnan Observatory, CAS, for their unparalleled hospitality during visits
when this project was developed.  His participation was supported in part by US National Science
Foundation grant AST 04-06726.  And finally, we thank an anonymous referee for constructive coments
and criticisms to improve the clarity of this paper.

\end{document}